\documentclass[fleqn,usenatbib,useAMS]{mnras}

\usepackage{newtxtext,newtxmath}
\usepackage[utf8x]{inputenc}
\usepackage{latexsym}
\usepackage[T1]{fontenc}
\usepackage{ae,aecompl}
\usepackage{float}
\usepackage{graphicx}
\usepackage{amsmath,amsfonts}
\usepackage{lscape}
\usepackage{afterpage}
\usepackage[table]{xcolor}
\usepackage{threeparttable}
\usepackage{mathtools}
\usepackage{pdflscape}
\usepackage[export]{adjustbox}% http://ctan.org/pkg/adjustbox

\title[VLA observations towards the ONC]{From downtown to the outskirts: a radio survey of the Orion Nebula Cluster}

\author[J. Vargas-González et al.]
{
J. Vargas-Gonz\'{a}lez,$^1$\thanks{E-mail: j.i.vargas-gonzalez@herts.ac.uk}
J. Forbrich,$^{1,2}$
S. A. Dzib,$^3$
and J. Bally, $^4$
\\
$^1$Centre for Astrophysics Research, School of Physics, Astronomy and Mathematics, University of Hertfordshire, College Lane, Hatfield AL10 9AB, UK\\
$^2$Center for Astrophysics | Harvard \& Smithsonian, 60 Garden St, Cambridge MA 02138, USA\\
$^3$Max-Planck-Institut für Radioastronomie, Auf dem Hügel 69, 53121 Bonn, Germany\\
$^4$Center for Astrophysics and Space Astronomy, Astrophysical and Planetary Sciences Department, University of Colorado, UCB 389 Boulder, Colorado 80309, USA
}

\date{Accepted XXX. Received YYY; in original form ZZZ}

\pubyear{2021}

\begin{document}
\label{firstpage}
\pagerange{\pageref{firstpage}--\pageref{lastpage}}
\maketitle

%%%%%%%%%%%%%%%%%%%%%%%%%%%%%%%%%%%%%%%%%%%%%%%%%%%%%%%%%%%%%%%%
%%%%%%%%%%%%%%%%%%%        ABSTRACT       %%%%%%%%%%%%%%%%%%%%%%
%%%%%%%%%%%%%%%%%%%%%%%%%%%%%%%%%%%%%%%%%%%%%%%%%%%%%%%%%%%%%%%%
\begin{abstract}
We present a newly enlarged census of the compact radio population towards the Orion Nebula Cluster (ONC) using high-sensitivity continuum maps (3-10 $\mu$Jy bm$^{-1}$) from a total of $\sim30$~h centimeter-wavelength observations over an area of $\sim$20$\arcmin\times20\arcmin$ obtained in the C-band (4$-$8~GHz) with the Karl G. Jansky Very Large Array (VLA) in its high-resolution A-configuration. We thus complement our previous deep survey of the innermost areas of the ONC, now covering the field of view of the Chandra Orion Ultra-deep Project (COUP). Our catalog contains 521 compact radio sources of which 198 are new detections. Overall, we find that 17\% of the (mostly stellar) COUP sources have radio counterparts, while 53\% of the radio sources have COUP counterparts. Most notably, the radio detection fraction of X-ray sources is higher in the inner cluster and almost constant for $r>3\arcmin$ (0.36~pc) from $\theta^1$ Ori C suggesting a correlation between the radio emission mechanism of these sources and their distance from the most massive stars at the center of the cluster, for example due to increased photoionisation of circumstellar disks. The combination with our previous observations four years prior lead to the discovery of fast proper motions of up to $\sim$373~km s$^{-1}$ from faint radio sources associated with ejecta of the OMC1 explosion. Finally, we search for strong radio variability. We found changes in flux density by a factor of $\lesssim$5 within our observations and a few sources with changes by a factor $>$10 on long timescales of a few years.
\end{abstract}


%%%%%%%%%%%%%%%%%%%%%%%%%%%%%%%%%%%%%%%%%%%%%%%%%%%%%%%%%%%%%%%%
%%%%%%%%%%%%%%%%%%%     INTRODUCTION      %%%%%%%%%%%%%%%%%%%%%%
%%%%%%%%%%%%%%%%%%%%%%%%%%%%%%%%%%%%%%%%%%%%%%%%%%%%%%%%%%%%%%%%
\section{Introduction}

The advent of wideband centimeter-wavelength observing capabilities has enabled a new era of stellar radio astronomy, including observations of radio counterparts of young stellar objects (YSOs). Generally, these show evidence for both thermal free-free emission from ionized material and also nonthermal (gyro-)synchrotron emission from magnetospheric activity \citep{dul85, gue02}. A few years ago, in \citet{for16}, we used the National Radio Astronomy Observatory's upgraded Very Large Array (VLA) to study radio counterparts of YSOs in the Orion Nebula Cluster (ONC), the richest nearby young stellar cluster. This study, based on pointing the VLA at the heart of the ONC for about 30 hours, increased the number of known radio sources in the ONC by a factor of $\sim$7 with the detection of 556 compact sources. Other studies, using the VLA toward the ONC have also focused in the inner cluster covering similar areas (within $\sim6\arcmin\times6\arcmin$) usually reaching rms noise levels above $30\;\mu$Jy bm$^{-1}$ limiting the number of detections from a few tens (prior to the VLA upgrade \citealt{chu87, gar87, fel93b, zap04}) to up to 175 sources \citep{she16}. In a larger surveyed area, \cite{kou14} obtained a shallow map of approximately $1\fdg6\times0\fdg4$ around the ONC reporting a total of 165 sources and typical rms noise levels of 60 $\mu$Jy bm$^{-1}$.

Additionally, multi-epoch VLA data with its high angular resolution ($0\farcs1$) and astrometric capabilites have been used to constrain the kinematics of the ONC \citep{gomez2005, kou14, dzib17}. The main focus has been on the improvement of proper motion (PM) estimates of the main stellar radio sources in the Orion Becklin–Neugebauer/Kleinmann–Low region (BN/KL) in the inner ONC \citep{becklin1967, kleinmann1967, gomez2008, zapata2009, rod17, rodriguez2020}. The PM of these stars supports the scenario where a multiple stellar system experienced a close dynamical interaction resulting in two massive stars (BN source and source I) and at least one low-mass star being ejected at a few tens of km s$^{-1}$ triggering a powerful outflow emerging from the OMC1 cloud core prominently seen at near-infrared wavelengths \citep{bally2015}. The improved sensitivity of the VLA now enables the detection also of non-stellar emission like that from jets and outflows \citep{for16, bally2020}, enabling astrometric studies as presented here.

While the deep radio catalog presented in \cite{for16} considerably improved the census of compact radio sources it left open the question of the wider radio population in the ONC, and its interplay with the well-characterized X-ray and infrared populations. The Chandra Orion Ultra-deep Project (COUP; \citealp{get05b}) covers a larger area ($\sim17\arcmin\times17\arcmin$) than our single, deep VLA pointing around the same reference center, and we thus conducted a wider survey for radio sources in the ONC which is presented in this work. Six additional pointings surrounding the deep central pointing were obtained at an unprecedented sensitivity in this area and we also repeated the central pointing for comparison (see Figure \ref{fig_obs_setup}).

The new observations discussed here allow us to obtain the largest census to date of radio counterparts to YSOs anywhere, which we can place into the rich multi-wavelength context of the ONC. We additionally make use of radio astrometry in a comparison of the two central pointings, separated by only $\sim$4.15 years, to study fast proper motions in the ONC, which is mainly of interest for non-stellar emission, which otherwise are more difficult to measure while providing valuable additional information for source identification. Finally, we use the excellent sensitivity even on short timescales to continue our study of YSO radio variability, motivated by the findings of extreme variability on short timescales (factor $>$138 in less than two days and a factor of 10 in less than 30 minutes) in our previous deep ONC pointing \citep{for17}.

%%%%%%%%%%%%%%%%%%%%%%%%%%%%%%%%%%%%%%%%%%%%%%%%%%%%%%%%%%%%%%%%
%%%%%%%%%%%%%%%%%%%     OBSERVATIONS      %%%%%%%%%%%%%%%%%%%%%%
%%%%%%%%%%%%%%%%%%%%%%%%%%%%%%%%%%%%%%%%%%%%%%%%%%%%%%%%%%%%%%%%
\section{Observations and Data Reduction}\label{obs}

\begin{figure*}
	\includegraphics[width=0.75\linewidth]{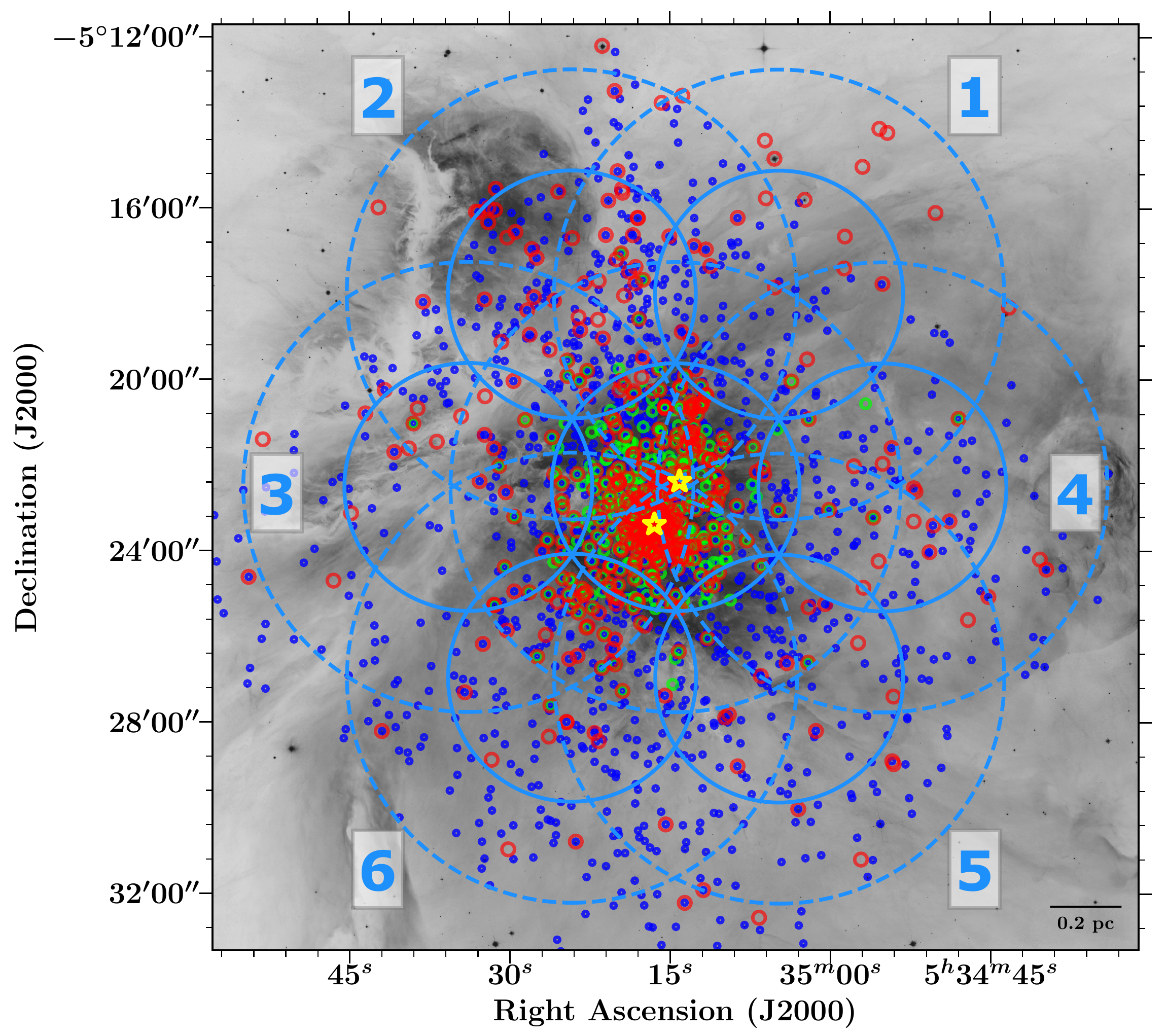}
    \caption{Observational setup: the central pointing, identical to the pointing 
    position of the observations presented by \citet{for16}, is surrounded by six 
    additional pointings (listed in Table \ref{obs_and_images}) with the same spectral 
    setup in C-band. The light-blue circles indicate the HPBW of each pointing at the 
    low (4288 MHz) and high (7847 MHz) frequency ends of the bandwidth in dashed and 
    continuous lines ($\sim 10\farcm5$ and $\sim 5\farcm8$, respectively). Red symbols 
    show the radio sources detected in this work, green symbols indicate radio sources 
    detected in the deep survey \citep{for16}, with yellow symbols additionally marking 
    the positions of $\theta^1$ Ori C and the BN object for reference (lower left and upper right, respectively). Blue symbols indicate 
    the positions of X-ray sources from the COUP survey \citep{get05a}. The background image 
    is a HST $r$-band image (ACS/WFC) of the Orion Nebula (Credit: NASA, ESA, M. Robberto, 
    and the Hubble Space Telescope Orion Treasury Project Team).}
    \label{fig_obs_setup}
\end{figure*}

The radio data were obtained between October and November 2016 using 
the NRAO\footnote{National Radio Astronomy Observatory is a facility 
of the National Science Foundation operated under cooperative agreement 
by Associated Universities, Inc.} VLA (project code: 16B-268). Figure 
\ref{fig_obs_setup} shows the observational setup which consists of a 
central pointing at $(\alpha$, $\delta)_{J2000}=(5^{\rm h}35^{\rm 
m}14^s_{\cdot}5$, $-5\degr22^{\rm m}30^s_{\cdot}6)$ and six adjacent 
pointings observed for about 4 hrs each with the most extended A-configuration 
array in C-band. The phase centers and dates of the observations are 
listed in Table \ref{obs_and_images}. The half-power beamwidth (HPBW) 
for the low-frequency limit (4288 MHz) is $\sim 10\farcm5$ covering a 
total area of $\sim20\times20$ arcmin within the collective low-frequency 
HPBW of all the pointings. The receivers were in full polarization mode 
with two basebands of 1 GHz each centred at 4.8 and 7.3 GHz with a total 
of 16 spectral windows (8 per baseband) divided into 64 channels of 2 MHz 
width each. The primary flux density calibrator for all the pointings was 
3C48 and the phase/gain calibrator was J0541-0541 observed every 5$-$6 min 
to ensure phase stability as in our earlier observations. 
The light-blue circles in Figure \ref{fig_obs_setup} represent the 
HPBW at the low (4288 MHz) and high (7847 MHz) frequency ends of 
the bandwidth in dashed and continuous lines ($\sim 10\farcm5$ 
and $\sim 5\farcm8$, respectively).

%%%%    Figure 1

%%%%   Table 1
\begin{table*}
\centering
\caption{VLA ONC Observations and main image parameters per pointing.}
\label{obs_and_images}
\begin{threeparttable}
\begin{tabular}{l c c c c c c}
\hline\hline
Pointing  & Starting Time &\multicolumn{2}{c}{Phase center}      & Synthesized beam size &  PA       &       RMS         \\
             &    (UTC)      & $\alpha_{2000}$ & $\delta_{2000}$ &      (FWHM)           & ($\degr$) & $\mu$Jy bm$^{-1}$  \\
\hline           
1       &  04 Oct 2016 / 08:42:27 & 5:35:04.7800 & -5:18:00.6700 &   $0\farcs33 \times 0\farcs28$    &  28  & $3.4$  \\
2       &  08 Oct 2016 / 11:40:20 & 5:35:24.1800 & -5:18:00.6700 &   $0\farcs35 \times 0\farcs24$    &  33  & $3.8$  \\
3       &  17 Oct 2016 / 09:51:32 & 5:35:33.8800 & -5:22:30.5700 &   $0\farcs34 \times 0\farcs24$    &  24  & $4.5$  \\
4       &  03 Oct 2016 / 09:35:25 & 5:34:55.1100 & -5:22:30.5700 &   $0\farcs35 \times 0\farcs26$    &  31  & $4.3$  \\
5       &  02 Oct 2016 / 12:53:50 & 5:35:04.7800 & -5:26:58.6201 &   $0\farcs51 \times 0\farcs22$    &  41  & $5.3$  \\
6       &  16 Oct 2016 / 11:14:20 & 5:35:24.1800 & -5:26:57.6899 &   $0\farcs38 \times 0\farcs24$    &  35  & $5.0$  \\
Centre  &  27 Nov 2016 / 04:49:20 & 5:35:14.4792 & -5:22:30.5760 &   $0\farcs36 \times 0\farcs25$    & -28  & $10.2$ \\
\hline
\end{tabular}
\centering
\begin{tablenotes}
\item Note: The light-blue number labels in Figure \ref{fig_obs_setup} correspond to the pointing numbers in this Table.
\end{tablenotes}
\end{threeparttable}
\end{table*}

The reduction of the data was performed using the VLA Calibration 
Pipeline using the CASA\footnote{Common Astronomy Software Application 
\citep{mcm07}.} software (release 5.4.1). All pointings except for pointing 
5 were reduced with the automatic processing of the pipeline. Pointing 
5 required additional manual selection of faulty data to be excluded 
in a small portion of the observation which includes 4 science scans 
and 1 calibrator scan (equivalent to a 5 min interval). No time or spectral averaging was applied to any of the different pointings.

The calibrated data were imaged with the TCLEAN task in CASA. All pointings 
were imaged to have a size of $8192\times8192$ pixels with a pixel size of 
$0\farcs1$ as a compromise between the pixel coverage of the synthesized beam 
and the final size of the image in order to cover the largest half-power 
beamwidth of $\sim$10$\farcm$5. We used the Stokes plane $I$ and spectral 
definition mode `mfs' (Multi-Frequency Synthesis) that combines the data 
from all the selected spectral channels into a single continuum image. The 
Hogbom deconvolution algorithm and a Briggs weighting method with a robustness 
parameter of 0.5  were used. For the central pointing, an additional set of images was created with similar 
setup but using the `mtmfs' algorithm (Multi-term Multi-Frequency Synthesis;  
\citealt{rau11}) with `nterms=2', generating Taylor-coefficient images 
corresponding to a continuum intensity and spectral index map. This imaging 
method used for the central pointing enables a direct comparison with the deep 
radio observations described in \citet{for16}. Spatial filtering of the visibility 
data was applied using baselines of the ($u$, $v$) range longer than 100 k$\lambda$ 
($\sim$6 km) to reduce the impact of extended nebular emission on the extraction of point-like 
sources by filtering out structures greater than $\sim2\arcsec$.

The main parameters of the resulting images for each pointing are listed 
in Table \ref{obs_and_images}, including the synthesized beam sizes and noise 
levels. The synthesized beam sizes are typically around $0\farcs3$
with only a slightly larger major axis of $\sim0\farcs5$ for pointing 5. The rms noise level, 
on the other hand, varies considerably throughout the different pointings 
reaching the highest value in the central pointing due to the complex 
structure of the inner part of the cluster. All the adjacent pointings 
have rms noise levels around $3-5\,\mu$Jy bm$^{-1}$ reaching the lowest 
value in pointing 1 in the north-west. 
The deep observation of the central pointing  
presented in \cite{for16} has a nominal rms noise of $3\,\mu$Jy bm$^{-1}$ 
being the most sensitive observations of the inner ONC to date at these 
frequencies. We reached similar rms noise in the outer pointings 
where most of the crowded area and complex structures lie towards the 
edges on these images. However, they still represent the most sensitive 
observations of the ONC to date while similar studies in this region have 
reported rms noise levels in the range of 25$-$80 $\mu$Jy bm$^{-1}$ 
\citep{zap04, kou14, she16}. 
To account for the radial decrease in sensitivity caused by the wideband primary beam response we have applied a primary beam correction factor to each source in our catalog after imaging the data following the method described in \citet{for16}. This primary beam correction factor is a function of distance to the phase center described by a polynomial.

For direct comparison with our earlier results and in order to determine proper motions, while widefield imaging within CASA is preliminary, we have used the standard gridder for our images. Our imaging experiments show that presently available widefield imaging (w-projection method through gridder='wproject') results in standard-gridder positions in the outer beam (r$>6\farcm$4) that can be slightly off by up to $0\farcs36$ from the corresponding wproject positions, i.e., up to the size of the synthesized beam, however there is only one source at such a large distance in our catalog presented in section \S \ref{source_detection}. At distances $r<3'$ from the phase center ($\sim$80\% of the sources in our catalog) the offsets found between these two gridders are negligible at $\lesssim 0\farcs04$ ($\lesssim$10\% the size of the synthesized beam). For even greater distances, the offset rises nonlinearly to about half the synthesized beam size ($0\farcs 2$) at $r=5\farcm 2$, encompassing 97\% of our sources. As discussed below, in our paper this has only a minor effect in catalog cross-matching for a limited number of sources in our catalog.

The imaging applied in both VLA epochs is identical, except that for the new epoch we have not applied any channel averaging, which is different from the approach in \citet{for16}. In order to quantify any impact from using these two different approaches we re-imaged the central pointing of the new observations with the same channel averaging used before. The positions in both images are compatible within the uncertainties where even at a large distance from the phase center (r=6$'$) the effect corresponds to a shift of less than 0.5$\sigma$ ($\sim$22~mas). The impact on flux densities is equally negligible.

%%%%%%%%%%%%%%%%%%%%%%%%%%%%%%%%%%%%%%%%%%%%%%%%%%%%%%%%%%%%%%%%
%%%%%%%%%%%%%%%%%%%        RESULTS        %%%%%%%%%%%%%%%%%%%%%%
%%%%%%%%%%%%%%%%%%%%%%%%%%%%%%%%%%%%%%%%%%%%%%%%%%%%%%%%%%%%%%%%
\section{Results}\label{results}

%%%%%%%%%%%%%%%%%%%          SOURCE DETECTION DISTRIBUTION          %%%%%%%%%%%%%%%%%%%%%%
\subsection{Source Detection and Distribution}\label{source_detection}

The high resolution provided by the A-configuration array together with the additional 
spatial filtering of the visibility data applied in the imaging process 
largely mitigated the difficulties of disentangling the compact radio emission 
from the range of emission size scales over the area due to the complex 
structure of the Orion Nebula. However, as reported by \citet{for16}, the use of automated 
methods for point source extraction leads to a high fraction of spurious 
detection (up to $50\%$) on VLA images with the same observational setup 
towards the ONC. 
The source detection was thus performed by visual inspection 
of each individual pointing using the images generated in the multi-frequency synthesis (mfs) spectral mode. The initial source positions in this process were estimated with DAOFIND 
task in IRAF. 
This task computes the positions by estimating the point spread 
function of the source using an elliptical Gaussian approximation within a given area 
defined by a box around the source. This first list of detections per pointing 
corresponds to the input for a source extraction script which uses the IMFIT 
task in CASA to obtain the final positions, flux densities and additional 
statistical parameters per source. For each input source we used different 
box sizes for the fitting ranging from 8 to 30 pixels, applying different 
offsets from the center of the source to avoid contaminant emission coming 
from nearby sources. 
The different outputs for a given source from different box configurations were compared to finally select the measurement with lower uncertainties and measured positions closer to the actual peak pixel, thereby minimising the impact of nearby sources and of complex nebular emission. We also enforced a minimum box size of 10$\times$10 pixels for it to contain at least about ten synthesized beam areas for statistics.
To select reliable detections we used a S/N$>$5 as our detection limit. 
Compared with a criterion of S/N$>$3, our adopted limit conservatively accounts for non-Gaussian noise in the inner cluster, which leads to many more sources that would need to be rejected (e.g., compact nebular emission).
A 5$\sigma$ cutoff is also consistent with the most complete radio surveys to date in the ONC field with detection limits between $4.5-6\sigma$ \citep{kou14, she16, for16}.

%%%%   Table 2
\begin{table}
\centering
\caption{Source detection per pointing.}
\label{source_detection_table}
\begin{threeparttable}
\begin{tabular}{c c r r r}
\hline\hline
Pointing ID & \multicolumn{2}{c}{Number of detections}    & \multicolumn{2}{c}{New detections}                              \\
            & Total &\multicolumn{1}{c}{$<$HPBW\tnote{a}} & \multicolumn{1}{c}{Total} & \multicolumn{1}{c}{$<$HPBW\tnote{a}}\\
    (1)     & (2)   &\multicolumn{1}{c}{(3)}              & \multicolumn{1}{c}{(4)}   & \multicolumn{1}{c}{(5)}             \\
\hline   
1       & 176 &16\ \ \ (9\%) & 44 (25\%)     & 14\ \ \ (8\%)\\
2       & 214 &  54 (25\%)   & 72 (34\%)     & 43 (20\%)\\
3       & 228 &  31 (14\%)   & 41 (18\%)     & 16\ \ \ (7\%)\\
4       & 158 &15\ \ \ (9\%) & 24 (15\%)     &  9\ \ \ (6\%)\\
5       & 151 &  19 (13\%)   & 26 (17\%)     & 10\ \ \ (7\%)\\
6       & 224 &  56 (25\%)   & 35 (16\%)     & 19\ \ \ (8\%)\\
Central & 272 & 249 (92\%)   & 22\ \ \ (8\%) & 19\ \ \ (7\%)\\
\hline
\end{tabular}
\centering
\begin{tablenotes}
\item[a] HPBW at the high frequency end is $5\farcm8$ ($r\sim2\farcm9$ from the phase center).
\item All the percentages are with respect to column (2).
\end{tablenotes}
\end{threeparttable}
\end{table}

The final number of sources detected per pointing is listed  in column 
(2) in Table \ref{source_detection_table} and marked in red in Figure 
\ref{fig_obs_setup}. Column (3) in the table indicates the number of 
sources within the HPBW (at the high frequency end $\sim5\farcm8$) 
for each pointing. The maximum number of detections is found in the 
central pointing with 272 detections (which corresponds to the densest 
region of the ONC) and the lowest detection count is found in pointing 5. 

The total number of detections over the whole area covered by the 7 
pointings is 521, without duplicates that are detected in more than 
one pointing. Since most of the duplicates are isolated sources, those 
were easily found with a search radius of 1", followed by visual inspection.

%%%%%%%%%%%%%%%%%%%%%%%%%%%%%%%%%%%%%%%%%%%%%%%%%%%%%%%%%%%%%%%%%%%
%%%%%%%%%%%%%%%%%. Position shifts discussion. %%%%%%%%%%%%%%%%%%%%
%%%%%%%%%%%%%%%%%%%%%%%%%%%%%%%%%%%%%%%%%%%%%%%%%%%%%%%%%%%%%%%%%%%

Sources detected in several pointings are usually found at different distances from the phase centers. As noted above, these different detections show minor position shifts. In our catalog we thus report positions from the closest detection to the phase center, where this corresponding distance is also reported.
Partly due to the spatial filtering inherent in our experiment design, focusing on the detection of nonthermal emission with aggressive spatial filtering, most of the sources are unresolved, with only 2\% showing a size $\geq$3 times the area of the synthesized beam, and only 8\% of the sources larger than twice the synthesized beam. The nominal mean ratio between the integrated to peak flux density in our catalog is $1.2\pm0.5$. Our catalog thus only lists peak flux densities.

Our catalog of 521 sources is listed in Table \ref{catalog_tab} and indicated by red symbols in Figure \ref{fig_obs_setup}. Columns (1) and (2) show the positions in $\alpha$ and $\delta$ with their corresponding uncertainties obtained from the fit (IMFIT). Column (3) indicates the source identification number in this catalog. Columns (4) and (5) indicate the peak flux density (corrected by the primary beam response) and source fitting parameters (major and minor axes, and position angle). Columns (6) and (7) indicate previous designation in the COUP and/or VISION surveys. Column (8) indicates the distance to the stellar system $\theta^1$ Ori C and column (9) indicates the distance to the closest phase center where the reported positions come from. The position uncertainties reported in this catalog are those given by IMFIT without the addition of minor systematic errors (see above)\footnote{We have applied two different and complementary cutoff methods to discuss the proper motion significance in Section \S\ref{PM_section}.}. These positional uncertainties have median values of 14 and 15 mas in R.A. and decl., respectively. An absolute astrometric accuracy for similar VLA observations (identical to our central pointing) was reported in \cite{for16} using five individual epochs resulting in an overall absolute astrometric accuracy of 20-30 mas. An absolute uncertainty in peak flux densities of 5\% has been estimated based on systematic variability using a non-variable test-case (source BN; see section \S \ref{variability}). This uncertainty has been added in quadrature with the uncertainties from the 2D-Gaussian fit (IMFIT) already corrected by the primary beam response.

%%%%   Table 3
\begin{table*}
\centering
%\small
\caption{Catalog of compact radio sources in the ONC.}
\label{catalog_tab}
\begin{threeparttable}
\tabcolsep=0.11cm
\begin{tabular}{ccccccccc}
\hline \hline
$\alpha(2000)$      & $\delta(2000)$             & ID &Peak Flux Density& Deconvolved Size $^a$                   & COUP & VISION & Dist. to $\theta^1$ Ori C & rad $^b$\\
$({\rm ^h\,^m\,^s})$& $(\degr\,\arcmin\,\arcsec)$&    &(mJy bm$^{-1}$)&$(\theta_{max}\times\theta_{min}$ ; PA) & & & (arcmin) & (arcmin) \\
(1) & (2)  &  (3)  & (4)  & (5) & (6)  &  (7)  &  (8) & (9)\\
\hline
05:34:39.7603 $\pm$ 0.0002 &-5:24:25.465 $\pm$ 0.003 & 1  & 1.422 $\pm$ 0.074 & $0\farcs27\times0\farcs17\,;\,45\degr\pm5\degr$  &    & 05343976-0524254 & 9.2  &4.3\\
05:34:40.3667 $\pm$ 0.0006 &-5:24:11.308 $\pm$ 0.009 & 2  & 0.148 $\pm$ 0.011 & $0\farcs21\times0\farcs06\,;\,64\degr\pm20\degr$ &    &                  & 9.0  &4.0\\
05:34:43.2598 $\pm$ 0.0028 &-5:18:18.566 $\pm$ 0.020 & 3  & 0.045 $\pm$ 0.009 &                                                  &    &                  & 9.7  &5.1\\
05:34:45.1880 $\pm$ 0.0019 &-5:25:03.941 $\pm$ 0.016 & 4  & 0.047 $\pm$ 0.006 &                                                  & 23 & 05344519-0525041 & 8.0  &3.6\\
05:34:47.0976 $\pm$ 0.0007 &-5:25:36.158 $\pm$ 0.017 & 5  & 0.063 $\pm$ 0.007 &                                                  &    & 05344709-0525363 & 7.6  &3.7\\
05:34:47.9813 $\pm$ 0.0001 &-5:20:54.381 $\pm$ 0.002 & 6  & 0.406 $\pm$ 0.021 & $0\farcs16\times0\farcs12\,;\,117\degr\pm18\degr$ &   &                  & 7.5  &2.4\\
05:34:48.8288 $\pm$ 0.0002 &-5:23:17.906 $\pm$ 0.003 & 7  & 0.180 $\pm$ 0.010 & $0\farcs09\times0\farcs02\,;\,49\degr\pm64\degr$ & 43 & 05344883-0523179 & 6.9  &1.8\\
05:34:50.1332 $\pm$ 0.0038 &-5:16:06.062 $\pm$ 0.020 & 8  & 0.040 $\pm$ 0.007 &                                                  &    &                  & 9.8  &4.1\\
05:34:50.3353 $\pm$ 0.0013 &-5:23:23.775 $\pm$ 0.028 & 9  & 0.019 $\pm$ 0.003 &                                                  &    &                  & 6.5  &1.5\\
05:34:50.7098 $\pm$ 0.0023 &-5:24:01.184 $\pm$ 0.016 & 10 & 0.020 $\pm$ 0.004 &                                                  & 57 & 05345071-0524013 & 6.4  &1.9\\
05:34:52.0114 $\pm$ 0.0002 &-5:22:36.387 $\pm$ 0.006 & 11 & 0.088 $\pm$ 0.005 &                                                  &    & 05345201-0522364 & 6.1  &0.8\\
05:34:52.1746 $\pm$ 0.0007 &-5:22:31.786 $\pm$ 0.017 & 12 & 0.027 $\pm$ 0.003 &                                                  & 67 & 05345216-0522319 & 6.1  &0.7\\
05:34:52.1839 $\pm$ 0.0011 &-5:23:18.237 $\pm$ 0.027 & 13 & 0.013 $\pm$ 0.002 &                                                  &    &                  & 6.0  &1.1\\
05:34:54.0616 $\pm$ 0.0025 &-5:28:58.287 $\pm$ 0.049 & 14 & 0.038 $\pm$ 0.006 &                                                  &    &                  & 7.9  &3.3\\
05:34:54.0843 $\pm$ 0.0034 &-5:27:23.612 $\pm$ 0.059 & 15 & 0.042 $\pm$ 0.007 &                                                  &    &                  & 6.9  &2.7\\
05:34:54.1948 $\pm$ 0.0022 &-5:28:54.243 $\pm$ 0.030 & 16 & 0.069 $\pm$ 0.007 &                                                  & 90 & 05345419-0528543 & 7.8  &3.3\\
05:34:54.2494 $\pm$ 0.0003 &-5:21:35.423 $\pm$ 0.006 & 17 & 0.068 $\pm$ 0.005 &                                                  & 89 & 05345425-0521354 & 5.8  &0.9\\
05:34:54.6354 $\pm$ 0.0012 &-5:14:13.927 $\pm$ 0.023 & 18 & 0.042 $\pm$ 0.007 &                                                  &    &                  & 10.6 &4.6\\
05:34:55.0949 $\pm$ 0.0006 &-5:21:57.572 $\pm$ 0.010 & 19 & 0.041 $\pm$ 0.004 &                                                  &    &                  & 5.5  &0.6\\
05:34:55.0981 $\pm$ 0.0005 &-5:17:45.489 $\pm$ 0.009 & 20 & 0.076 $\pm$ 0.006 & $0\farcs22\times0\farcs16\,;\,28\degr\pm65\degr$ &    &                  & 7.7  &2.4\\
\hline
\end{tabular}
\centering
\begin{tablenotes}
\item $^a$ As defined by imfit in CASA.
\item $^b$ Distance to closest phase center. Note that the reported positions are from the standard gridder and may be slightly off in the outermost beam areas (see text).
\item The full catalog is available as supplementary material.
\end{tablenotes}
\end{threeparttable}
\end{table*}

In this new census of compact radio 
sources towards the ONC we report 198 new sources not previously reported 
at these frequencies. This is based on a comparison against the most 
complete catalogs of compact radio sources made of the ONC and presented 
in \cite{for16, she16, kou14, zap04}. The highest fractions of new sources 
are found in pointings 1 and 2, the northernmost part of the cluster as 
seen in Figure \ref{fig_obs_setup}. 

For a more direct comparison the central pointing was compared against the 
deep observation presented in \cite{for16} which are identical observations 
apart from the cumulative observing time ($\sim$30 h for the 
deep obs.). In the same area where they find 556 sources we detect 272. There 
are 303 sources in the deep catalog not detected in the central pointing of 
the new observations (not all the 272 have a counterpart in the deep catalog) 
which is expected due to the difference in sensitivity, where a large fraction 
of faint sources in the deep catalog are far below the noise level in the new 
data. However, not all the sources detected in 
the central pointing of the new and less sensitive observations were detected 
in the deep data, and indeed there are 19 new sources in the central pointing 
that should have been clearly detected in the deep data. These are clear targets 
for variability analysis and are discussed in section \S \ref{variability}.

\subsection{Multi-wavelength Populations}\label{xray_ir_counterpart}

Our primary goal after revealing the compact radio population in the ONC is the search for radio counterparts to YSOs. They can be detected through both thermal and nonthermal radio emission. While nonthermal emission originates at the smallest scales in the stellar coronae of low-mass young stars, thermal radio emission, in contrast, occurs in a range of larger scales in stellar disks or outflows when it is associated to a stellar source. However, compact sources at these frequencies can also be detected as thermal emission from ionized material not strictly related to a stellar source \citep{gar87, chu87, she16, for16}. An additional source of contamination comes from extragalactic background sources, even though these have been previously found to be minimal. In the so far deepest catalog for the inner ONC \citep{for16} it was estimated that $\sim$97\% of the compact radio sources within $r<1\farcm6$ are related to the cluster. Here we cover a considerably wider area and while the central region is less prone to present background detections due to the elevated rms noise caused by the nebula itself, we expect to find faint background radio sources from bright background galaxies in the outer areas, given the high sensitivity. A discussion on the expected number of extragalactic radio sources in the outer areas is presented later on in this section, when discussing Figure \ref{radio_no_xray}. 

In order to approach the nature of the compact radio population we have compared our catalog against the most complete X-ray and near-infrared (NIR) datasets available for the ONC. The COUP X-ray catalog reported in \cite{get05b} represents the best reference for YSOs in the ONC. Here, the X-ray emission of young stars is associated to thermal emission from hot plasma in coronal-type activity (e.g., \citealp{fem99}). The observations presented in this work were designed to cover the COUP survey area of $\sim$20$\arcmin \times 20\arcmin$ (blue markers in Figure \ref{fig_obs_setup} represent the COUP catalog). The complementary NIR dataset used here is the VISION catalog reported in \cite{mei16}, a survey of the entire Orion A molecular cloud in $JHK_S$ bands. The NIR band is an additional tracer of the young stellar population towards the ONC, since, in general, such X-ray and NIR data are similarly effective at detecting embedded sources (e.g., \citealp{ryt96}). Any extragalactic background contamination is unlikely where sources are superimposed onto the high extinctions levels of the ONC but at the same time the NIR also picks up foreground sources. Contrary to the X-ray survey, the NIR catalog is severely affected by bright extended emission of the Orion Nebula in the innermost areas, where it is thus less complete to YSOs. 

The angular resolution in COUP and VISION is $<0\farcs 5$, which is
comparable to the typical beam size in our observations. Moreover, the reported positional errors in the COUP survey have a mean value of $0\farcs1$, with maximum errors of $0\farcs59$, although, 99\% of the sources have positional errors $\leq0\farcs5$. On the other hand, the reported astrometric accuracy in the VISION catalogue is $\sim$70~mas. Based on these considerations and to additionally account for the smallest nearest-neighbour distances, we have therefore considered a conservative radius of $0\farcs5$ for the search of counterparts in these surveys. 
The robustness of this approach is demonstrated when considering that if we increase this search radius to $1''$ the number of correlations with the COUP and VISION surveys would increase by only 5\% and 4\%, respectively, including unrelated nearest neighbours.
Within the area covered in our observations we report 521 radio sources, while the COUP catalog reports 1616 sources and the VISION catalog 3558 sources. A total of 275  sources in our catalog ($\sim$53\%) have X-ray counterparts which corresponds to only $17\%$ of the COUP catalog. On the other hand, 290 NIR sources (or $\sim$56\%) have counterparts in our radio catalog, which corresponds to only $\sim$8\% of the NIR catalog within the same area.

%%%%   Figure 2
\begin{figure}
    \centering
    \includegraphics[width=\linewidth]{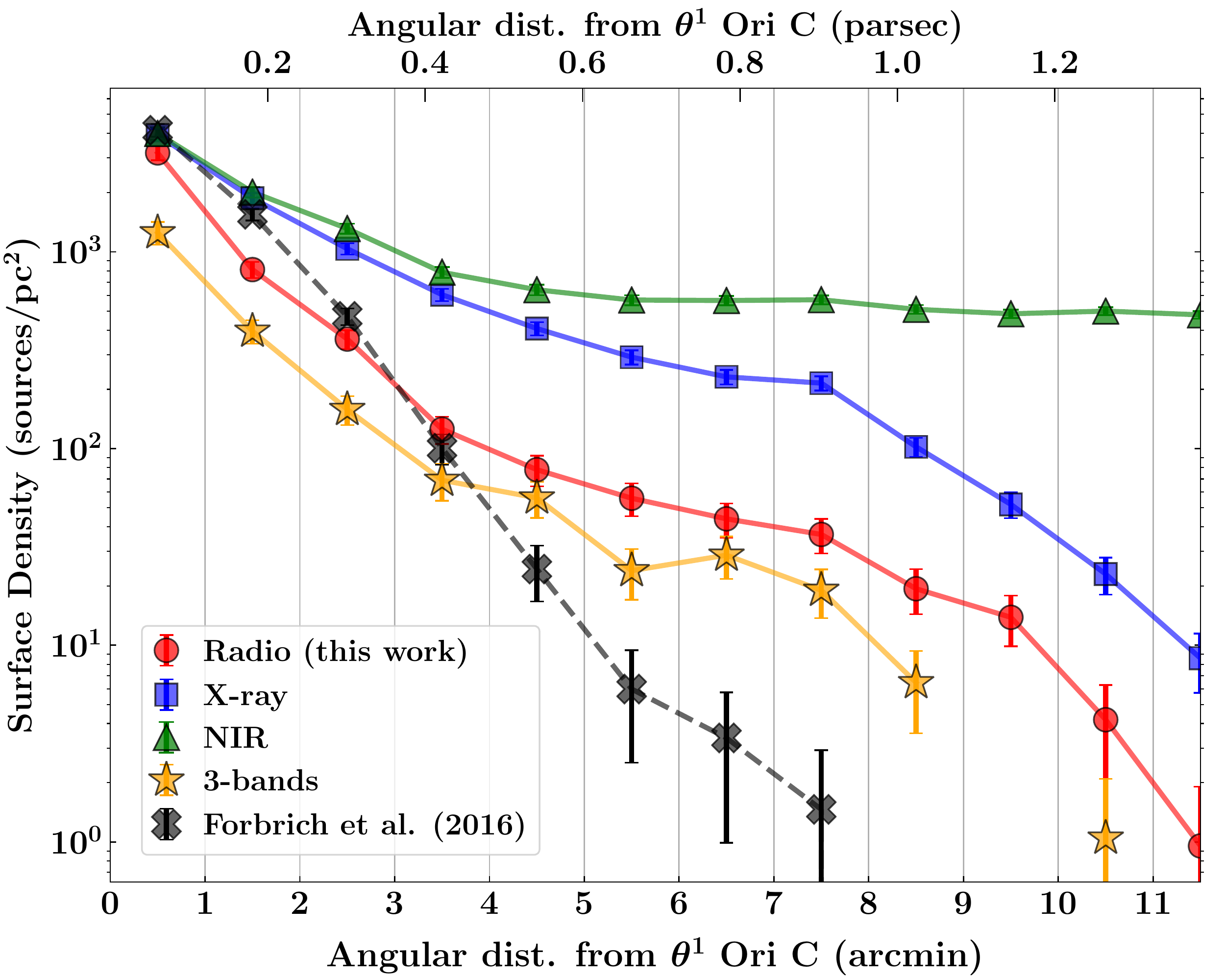}
    \caption{Azimuthally averaged surface number density as a function of projected distance to $\theta^1$ Ori C for the three different wavelength populations. The VISION, COUP and radio catalogs are indicated by green, blue and red lines, respectively. Sources with detections in all three bands are indicated in orange line, while the dashed black lines indicate the distributions for the deep catalog from \citet{for16}. Each data point represent the surface density of sources within annular areas of $1\arcmin$ width indicated by vertical lines and the error bars are based on error propagation from counting statistics (Poisson errors).}
    \label{master}
\end{figure}

Figure \ref{master} shows the surface number density of the three wavelength populations as a function of projected distance from $\theta^1$~Ori~C, a young and massive stellar system of $\sim$50 $M_\odot$ in the Trapezium cluster at the center of the ONC \citep{krauss2007}. Our observational setup is actually centered $\sim$1$\arcmin$ northwest from $\theta^1$~Ori~C towards the BN/KL region to optimize the sensitivity in this complex area while still matching the COUP field of view. Centering our reference on $\theta^1$ Ori C allows us to address any impact that the most massive stars in the center of the cluster have on the radio emission in a multiwavelength context. The overall distributions for the three bands show the densest region in the central cluster ($r<1\arcmin$) with a remarkable correspondence in their total number of sources equivalent to an average surface density of $(3.7\pm0.2)\times10^3$ sources pc$^{-2}$. Despite of this similarity, they do not represent the same population and, indeed, at this inner bin their actual correlation, indicated by the orange distribution (population with detections in the three bands), is only one third of their total number of sources. The X-ray and NIR distributions exhibit almost the same profile until $r=2\arcmin$. 

The X-ray distribution continuously decreases revealing, in part, the structure of the cluster but also showing the sensitivity limitations in the outer bins. Similarly, the radio distribution continuously decreases, although at a faster rate. Contrary to the deep radio catalog distribution (dashed black line), our radio catalog (continuous red line) does not decrease solely due to a sensitivity effect, even though the sensitivity is not constant. The angular distance between the phase centers of adjacent pointings is $\sim5\arcmin$ where the lowest sensitivity occurs at $2\farcm5$ between each pointing. At this distance, the wideband primary beam correction indicates a flux density correction by 40\%, which therefore technically constitutes the maximum variation in sensitivity due to the spacing of our individual pointings. However, we have shown in \cite{for16} that at this angular distance from the phase center of the central pointing the image noise is still dominated by the Orion Nebula itself, and the impact on our analysis is thus limited. Under this considerations, the radio distribution seems to reveal the intrinsic structure of the cluster. At larger radii ($r>4\arcmin$) the NIR distribution remains almost constant largely due to the presence of foreground infrared sources in addition to the young stellar population further away from the center of the cluster. Although the COUP distribution may also include contamination, in this case from extragalactic candidates, this is just a small fraction ($\sim$160) and most of the X-ray sources are likely members of the ONC \citep{get05b}, besides, extragalactic X-ray sources will be affected by foreground extinction in the cloud and thus are unevenly distributed.

%%%%   Figure 3
\begin{figure*}
    \centering
    \includegraphics[width=\linewidth]{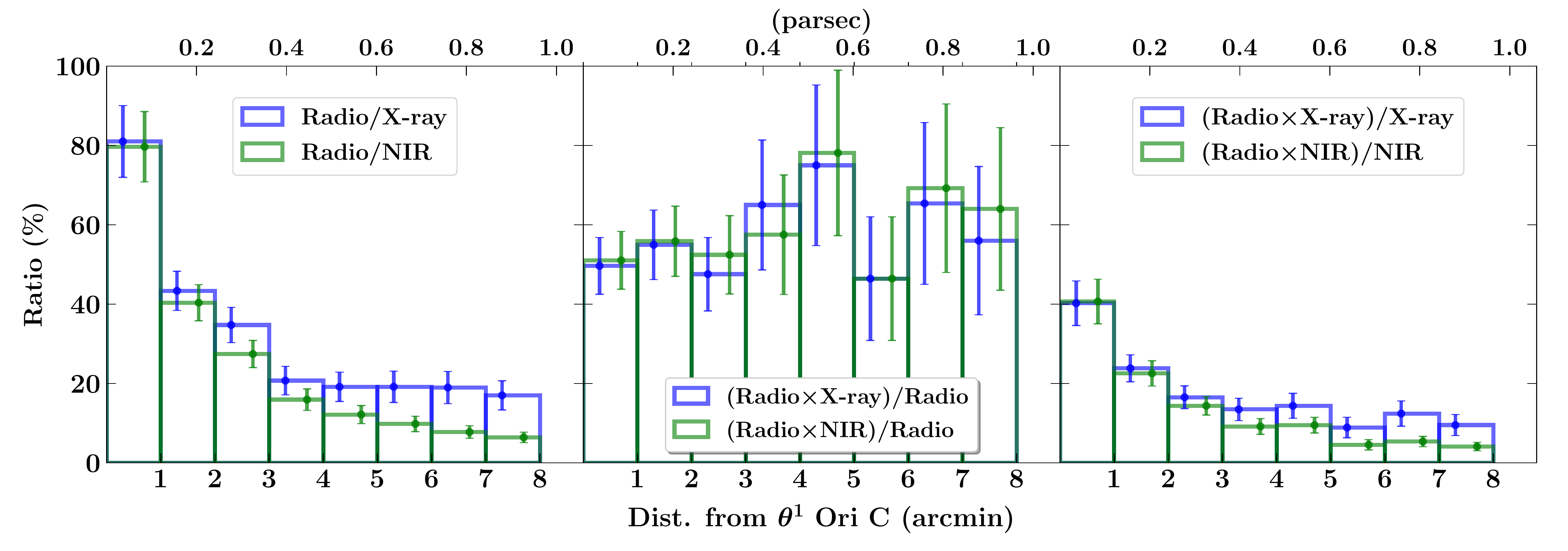}
    \caption{Detection fractions between the three different populations radio/X-ray/NIR as a function of distance to $\theta^1$ Ori C. The left panel shows the fraction of radio sources over the X-ray (blue) and NIR (green) populations. Central panel shows the X-ray (blue) and NIR (green) detection fraction of radio sources. The right panel shows the radio detection fraction of X-ray sources (blue) and NIR sources (green). The 1$\sigma$ error bars were derived from counting statistics (Poisson errors).}
    \label{det_frac}
\end{figure*}

In order to assess the correlation between the three main distributions shown in Figure \ref{master} (radio/X-ray/NIR) we quantified the different population fractions as a function of projected distance from $\theta^1$ Ori C as shown in Figure \ref{det_frac}.
Since we have shifted our reference center $\sim1\arcmin$ southeast from the actual center of the observations, 
we have therefore moved the reference frame for this analysis towards the south-east boundary of our radio survey (and COUP given the almost identical coverage). Since the maximum radial coverage in our radio catalogue is $\sim10\farcm4$, we are now reaching this boundary at $\sim9\farcm4$ south-east from $\theta^1$ Ori C, setting a first limit for our analysis at this maximum radius. However, an additional constraint is set to avoid any sensitivity bias for the detection of sources in the boundaries of our radio catalogue (which becomes sensitive to the brightest sources due to the primary beam correction). We have set a radial limit for our analysis to not exceed the midpoint between the HPBW at the low- and high-frequency ends of the bandwidth (dashed and continuous line circles, respectively, in Figure \ref{fig_obs_setup}). This midpoint is reached at $8'$ in the south-east direction, thus we are going to restrict our following analysis to this radius.
The left panel in Figure \ref{det_frac} indicates the fraction between the total radio distribution shown in Figure \ref{master} over the total X-ray and NIR distributions (indicated by blue and green histograms, respectively), without matching individual counterparts. In the central cluster these fractions indicate almost equal numbers of X-ray and NIR sources with radio source numbers lower by $\sim10-30\%$. These fractions then decrease as expected from their individual distributions in Figure \ref{master}. Here the radio to NIR fraction (green) decreases at a slightly faster rate due to the more uniform spatial distribution of NIR sources in the field (including foreground sources) compared to the continuously decreasing surface density distribution of radio sources as we go further away from $\theta^1$~Ori~C. Interestingly, for projected distances of $r>3\arcmin$, unlike the continuous decrease in the radio to NIR fractions, the radio to X-ray fractions (blue) settles at around $19\pm1\%$ which might indicate how these two populations are tracing a similar structure of the cluster within this radial range.

The middle panel in Figure \ref{det_frac} shows the X-ray and NIR detection fraction of radio sources (i.e., X-ray or NIR counterparts to our radio catalog) indicated by blue and green histograms, respectively. The overall distributions do not show a significant trend and they are, in effect, compatible with a constant distribution within the errors. Evidently, at larger radii the lower number of radio sources introduces larger uncertainties. A similar comparison is shown in the right panel, but indicating the radio detection fractions of X-ray and NIR sources (i.e., number of X-ray or NIR sources with radio counterparts) in blue and green, respectively. Here, the blue distribution represents our best estimate for the radio detection fraction of YSOs in the ONC and any radial trend with respect to the brightest Trapezium star at the center of the cluster may well provide important information on the underlying emission mechanism dominating at different distances. If YSOs have similar geometries, we would not expect to find any trend here, and the radio to X-ray ratio would instead simply reflect basic YSO properties. However, there is a very clear radial trend suggesting that X-ray sources towards the central part of the cluster are more likely to have a radio counterpart than X-ray sources in the outer areas. This points to a potential impact of the Trapezium on YSO properties in the ONC. One possibility is that circumstellar disks of YSOs (with a central X-ray source) are externally photoionized by the influence of the Trapezium stars (\citealt{odell1993, henney1998, con2020} and references therein), leading to the detection of ionized material as thermal free-free radio emission. In this case, as we go further away from the Trapezium, the detection fraction (blue distribution) may become dominated by nonthermal radio emission intrinsic to YSOs. The detection fraction is as high as $40\pm6\%$ in the inner bin ($r<0.12$ pc) and then decreases down to $17\pm3\%$ in the third bin ($r<0.36$ pc). For distances $r>3\arcmin$ (0.36 pc) the radio detection fraction of X-ray sources fluctuates around $11\pm2\%$ which might represent a baseline for the detection of nonthermal radio emission from the X-ray emitting YSO population without the influence of nearby young massive stars that would otherwise lead to a constant distribution. In contrast, the radio detection fraction of NIR sources (green) decreases continuously with radial distance, but this is expected since it reflects the fact that the NIR catalog includes not only cluster members but also sources that are not related to the ONC.

%%%%   Figure 4
\begin{figure}
    \centering
	\includegraphics[width=\linewidth]{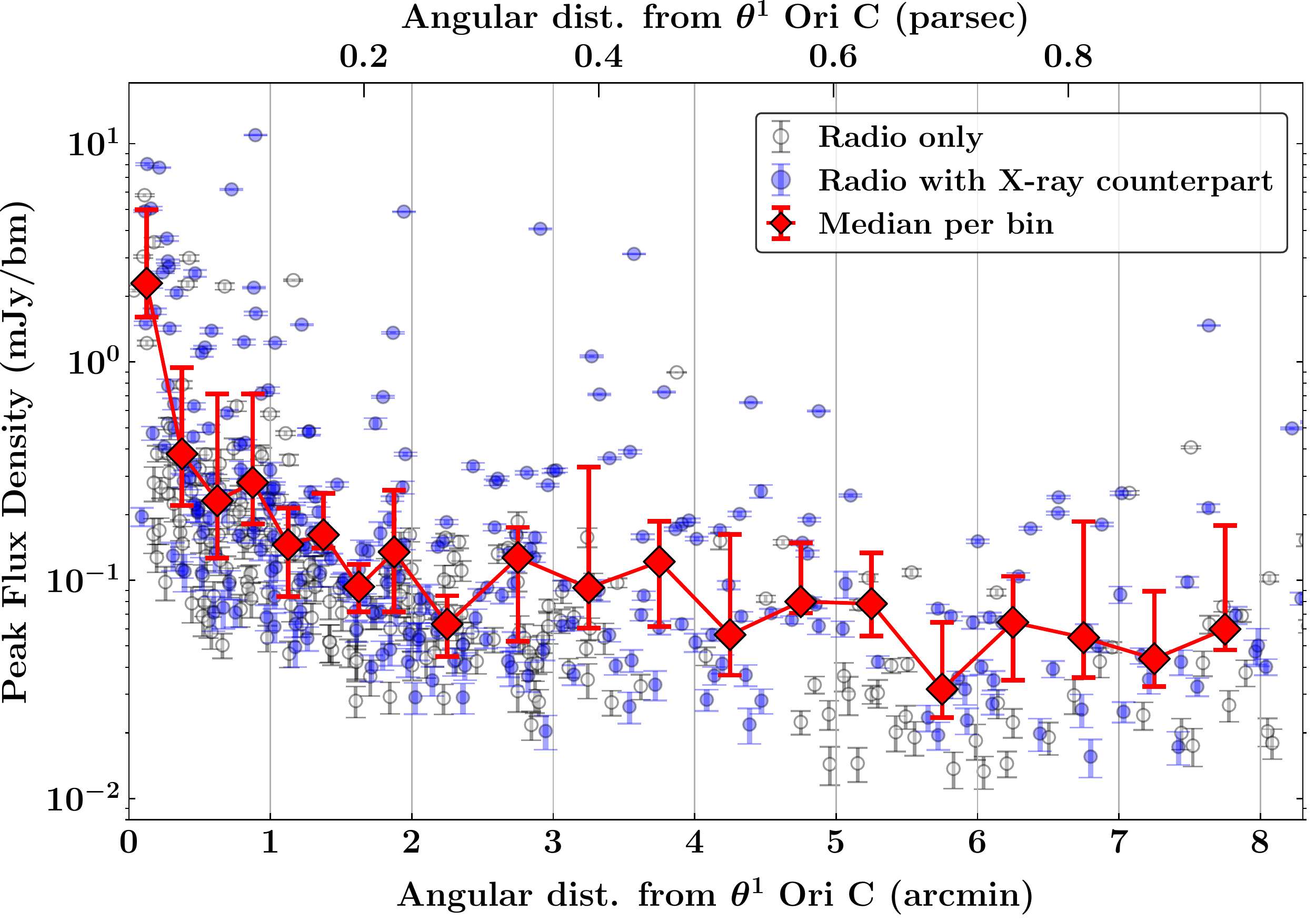}
    \caption{Peak flux density distribution as a function of projected distance to $\theta^1$ Ori C. Blue symbols indicate sources with X-ray counterparts and grey symbols indicate the remaining sources. Red symbols show the median of the blue distribution per $0\farcm25$ bins (for $r<2\arcmin$) and $0\farcm5$ bins (for $r>2\arcmin$). The median error bars represent the 25th and 75th percentiles of the data at each bin.}
    \label{flux_vs_dist}
 \end{figure}

Following the interpretation of the high radio detection fraction of X-ray sources closer to the Trapezium, more likely due to the detection of externally photoionized circumstellar disks, it is expected that the flux distribution of these sources (thermal free-free emission component) decreases with distance from the ionizing source. Figure \ref{flux_vs_dist} shows the peak flux density distribution as a function of projected distance to $\theta^1$ Ori C clearly showing higher radio fluxes closer to the Trapezium with a decreasing trend more evident for $r\lesssim2\farcm5$. Possible deviations from this trend are, first, the discrepancy between the projected and actual physical separation to the ionizing source where the actual physical separations could be considerably larger. Second, a fraction of the sources could be intrinsically bright nonthermal radio emitters.

%%%%   Figure 5
\begin{figure}
    \centering
    \includegraphics[width=\linewidth]{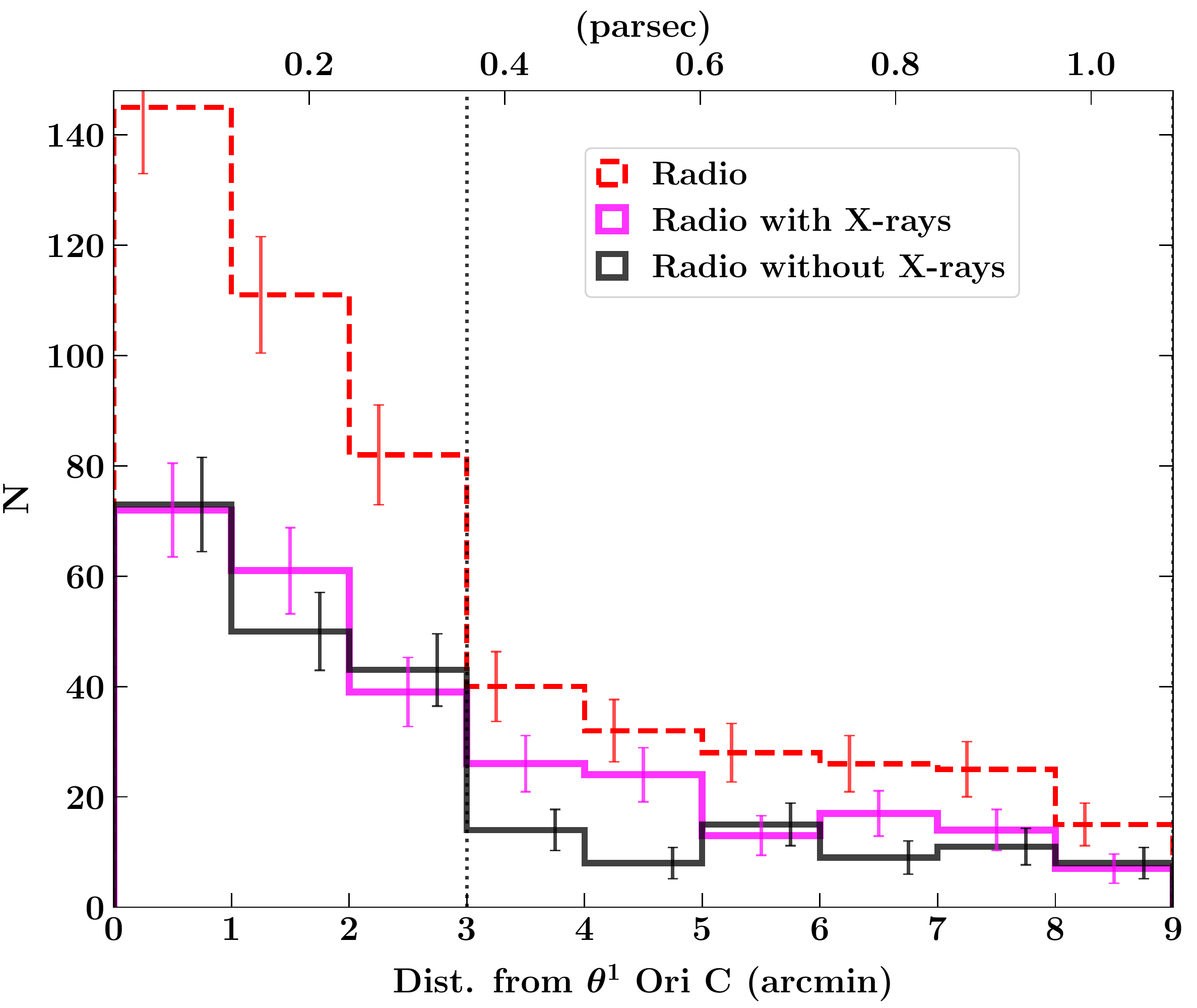}
    \includegraphics[width=\linewidth]{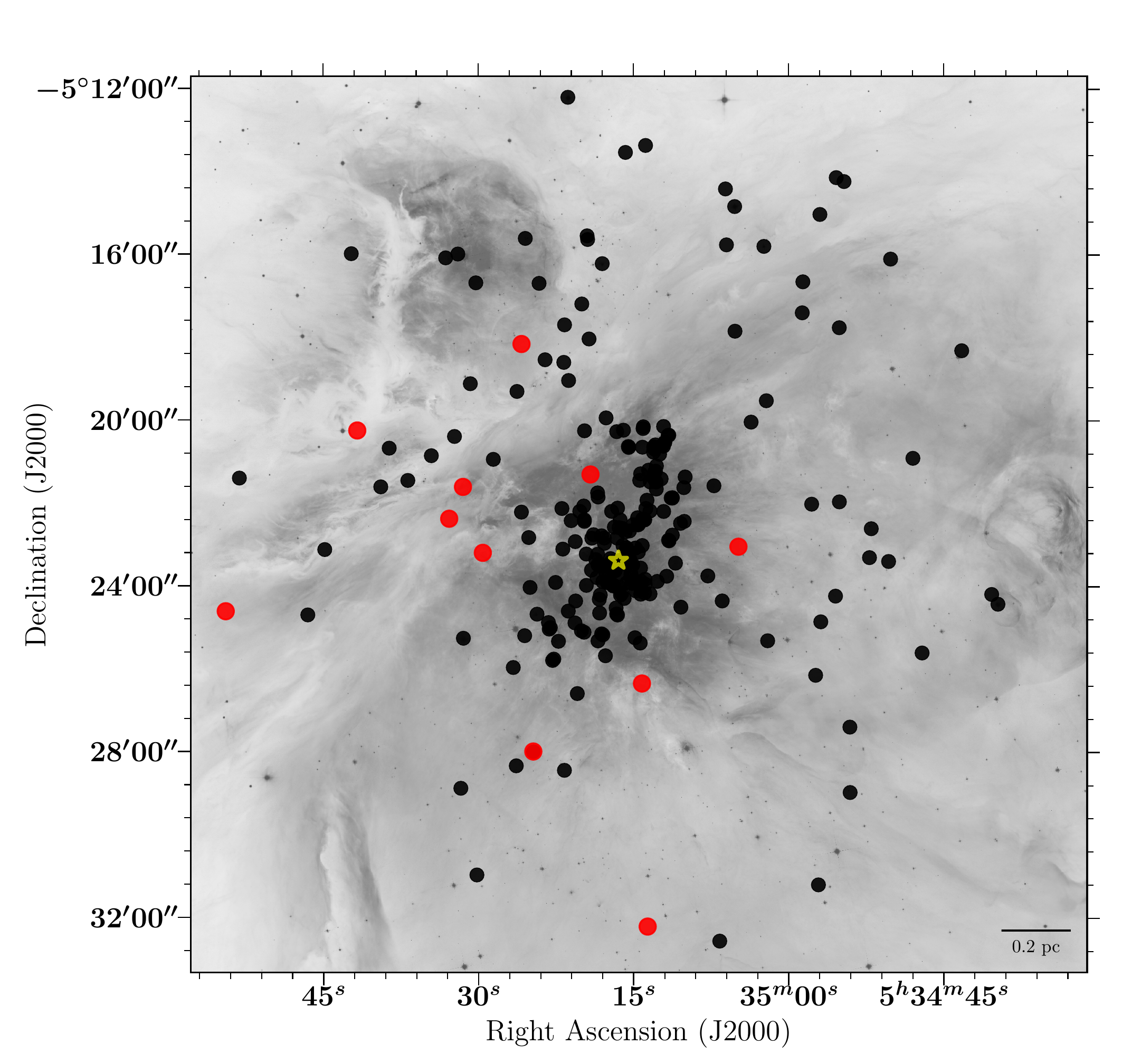}
    \caption{Top: Radial distribution of sources as a function of projected distance to $\theta^1$ Ori C. Red dashed line indicate the distribution of radio sources in our catalog. Magenta and black lines indicate the distribution of radio sources in our catalog with and without X-ray COUP counterparts, respectively. Vertical dotted line indicate the inner radius of the annular area used to estimate the expected number of extragalactic sources. Bottom: Spatial distribution of radio sources without X-ray counterpart (black) and radio sources with X-ray counterpart flagged as extragalactic candidates in the COUP catalog (red). The background image is a HST $r$-band image (ACS/WFC) of the Orion Nebula (Credit: NASA, ESA, M. Robberto, and the Hubble Space Telescope Orion Treasury Project Team).}
    \label{radio_no_xray}
\end{figure}

An interesting question prompted by these results concerns the nature of the radio population without X-ray counterparts. On the one hand, towards the center of the cluster most of these sources likely are thermal radio sources including externally photoionized disks, stellar jets and outflows, and also compact emission from the nebula, but towards the outer areas where we do not find a significant influence of the Trapezium stars we would expect a much lower number of these sources. As noted above, while the central area is less prone to background contamination already due to presence of the nebula, this contamination increases in the outer areas.
Additionally, there may be a low number of sources not identified as YSOs if these were not X-ray active at the time of the observations. 
Indeed, it has been found that X-ray sources with radio counterparts are predominantly sources with high X-ray luminosities and there are even a few cases where nonthermal radio YSOs have no X-ray counterpart at all \citep{fow13, riv15, for16}. The top panel in Figure \ref{radio_no_xray} shows the radial distribution of the total radio population, as well as the radio population with and without X-ray counterparts. The radio population without X-ray counterparts peaks in the central area, implying that there is a correlation with the inner cluster and these sources are not only background galaxies that would lead to a constant distribution as we see in the outer area ($>3\arcmin$). In this outer area the distribution of radio sources without X-ray counterparts is evenly spread over the field as indicated by black symbols in the bottom panel of Figure \ref{radio_no_xray}. In the same bottom panel we included in red symbols the radio sources with X-ray counterparts but flagged as extragalactic candidates in the COUP catalog. We found only 11 of these sources in our catalog ($\sim$2\%) out of 159 extragalactic source candidates in the COUP catalog, which are evenly spread throughout the whole field \citep{get05a}. 
In order to assess the possibility to detect extragalactic sources we made use of the work by \cite{fomalont1991} to estimate the expected number of background sources in a given area above a given flux density threshold. We considered an area between $3\arcmin<r<9\arcmin$ excluding the central cluster and where the distribution of radio sources without X-ray counterparts remains nearly constant around $11\pm3$ sources per bin, totaling $65\pm8$ sources. Following the 5$\sigma$ detection threshold used in our catalog and an average rms level of 4 $\mu$Jy bm$^{-1}$ (see Table \ref{obs_and_images}) we find an expected number of background sources of $\sim155\pm22$ and therefore extragalactic sources are a reasonable explanation for at least some of the radio sources without X-ray counterparts in the outer areas. In order to further explore the sample of 65 radio sources between $3\arcmin<r<9\arcmin$ we have looked into their correlations with NIR and optical wavelengths. 11 sources have NIR counterparts in VISION, 9 sources have optical counterparts in Gaia EDR3 \citep{gaia2016,gaia2021}, and just 4 sources have counterparts in both bands, however the nature of those remains unknown. On the other hand, 8 sources are associated with the OMC1 outflow discussed later in section \S \ref{PM_section}, and 4 sources are associated with proplyds found in optical HST data \citep{ric08}, leaving us with 55 unidentified sources in this sample and thus potentially extragalactic candidates.

We have focused our analysis on azimuthally averaged radial trends rather than individual pointings. This is justified by the fact that we find only limited evidence for significant differences between the individual pointings. In order to quantify any substantial differences between them we compared their relative multiwavelength distributions using KS tests. We found that the outer pointings have similar distributions and that the only large difference comes from comparing the central pointing with any of the outer ones (see section \S \ref{KS_test} for details on the KS statistics). 

Another interesting aspect is the relative fraction of radio sources that have neither X-rays nor NIR counterparts which is significantly lower in the outer fields. This population most likely represents compact radio emission of non-stellar origin. In section \S \ref{PM_section}, we identify one category of such non-stellar radio emission by employing proper motion measurements, identifying sources associated with the OMC1 outflow.

%%%%%%%%%%%%%%%%%%%                 PROPER MOTIONS                %%%%%%%%%%%%%%%%%%%%%%
\subsection{Proper Motions}\label{PM_section}

In our comparison between the observations presented here and the deep catalog reported in \cite{for16} we were able to implicitly measure the proper motion of 253 sources. This comparison only involves the central pointing in epoch 2016, these are identical observations towards exactly the same phase center and phase-referenced against the same distant quasar such that this comparison basically represents absolute astrometry. However, we are only covering a time baseline of 4.19 years. The mean position errors in our catalog are $\sim$20 mas and the estimated astrometric accuracy from the five epochs in \cite{for16} is 20-30 mas. In order to detect proper motions above these limits a single source would require a minimum velocity of $\sim$22~km~s$^{-1}$, which is unusual for a stellar source and therefore we did not expect to find significant motions in our sample. Stars with velocities $>$10~km s$^{-1}$ are considered peculiar sources and above $30$~km~s$^{-1}$ are already in the runaway regime \citep{farias2020, schoettler2020}. Stellar proper motions in the ONC are typically $\mu_{tot}\lesssim2$~mas~yr$^{-1}$ ($4$~km~s$^{-1}$) \citep{dzib17, kim2019} and only a handful number of sources have fast proper motions with $\mu_{tot}\leq$30~mas~yr$^{-1}$ (60~km s$^{-1}$) \citep{gomez2005, gomez2008, rodriguez2020} most of them associated to the ejected stars in the BN/KL region (which will be discussed later in this section).

\begin{figure*}
    \centering
	\includegraphics[width=0.8\linewidth]{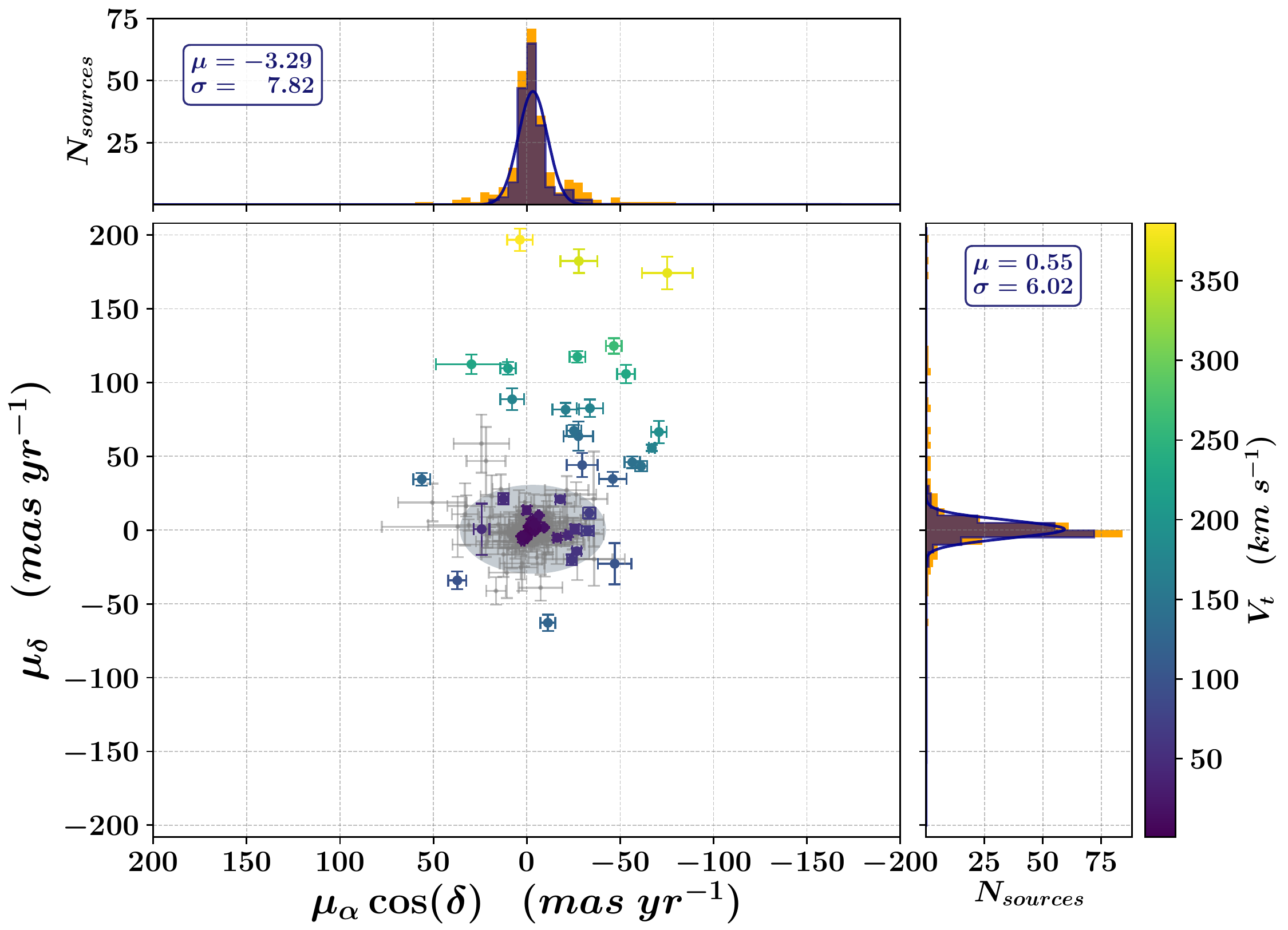}
    \caption{Absolute proper motions of compact radio sources in the $\mu_\alpha\cos(\delta)\ 
    \times\ \mu_\delta$ plane colour coded by their transverse velocity. Sources with motions 
    compatible with zero within 5$\sigma$ in $\alpha$ and $\delta$ are shown in grey. The histogram 
    distributions of motions in $\alpha$ and $\delta$ are shown in the top and right panel, 
    respectively, for the full sample (orange) and sub-sample (dark-blue). A Gaussian fit is 
    included for the sub-sample distribution used to define the intrinsic noise elliptical 
    region shown in the diagram.}
    \label{PM_full}
 \end{figure*}

In spite of the above considerations, we surprisingly discover high PMs in our sample with projected velocities of up to $\sim$373~km~s$^{-1}$ for a distance of $\sim400$ pc to the ONC \citep{grossschedl2018, kuhn2019}. This finding provides an additional and important tool for source identification. Figure \ref{PM_full} shows the absolute PM diagram in the $\mu_\alpha\cos(\delta)\ \times\ \mu_\delta$ plane for the full sample. Sources with motions above $5\sigma$ in at least one direction $\alpha$ or $\delta$ are colour coded by their transverse velocity ($V_t$). It is important to note here that these uncertainties were derived by error propagation from the fitting errors (positional uncertainties) thus these may be underestimated. 48 out of 253 sources fulfil this criterion. The remaining sources are marked in grey. With this first cutoff sources with small uncertainties but still with insignificant PM remain (even with motions above $5\sigma$). An additional cutoff is then applied taking into account the PM dispersion to estimate an intrinsic motion noise in the PM diagram in Figure \ref{PM_full}. This additional cutoff is described below.

An important caveat to consider in this sample is the fact that not all of the 
sources represent a stellar counterpart (see \S\ref{xray_ir_counterpart}) 
and therefore an unrealistic dispersion might be injected into the central part of 
the diagram caused by apparent motions of extended sources that can be seen as compact 
radio emission in form of thermal bremsstrahlung coming from ionized material around 
young stars, a peak emission in top of a bow shock, for instance. This apparent 
motion can be caused by fading and brightening events 
of the same feature in between the two epochs. On the other hand, given the short time 
separations between the two epochs we are only able to reveal significant motions 
above a few tens of mas yr$^{-1}$. However, there is a continuum range of motions 
spanning from $-$70 to 60~mas~yr$^{-1}$ in $\alpha$ and from $-$65 to 125~mas~yr$^{-1}$ 
in $\delta$. 

In order to determine a realistic PM dispersion we constrained a Gaussian 
fit for each axis in Figure \ref{PM_full} to a sample of sources that (1) are compact, (2) do not lie on a complex or extended emission and (3) do not present a clear motion which is considered as sources with position measurements compatible within 3$\sigma$ between the two epochs. This selection 
gives us a dispersion that represents the intrinsic motion noise in the PM diagram  
due to position uncertainties. This distribution is shown in dark-blue histograms 
in Figure \ref{PM_full} for each axis, together with their Gaussian fit also in 
dark-blue colour. The full sample distribution is showed in orange histograms. The 
shaded ellipse represent 5 times the dispersion of the Gaussian fit for each axis 
in order to exclude ambiguous motions within this area and focus the discussion 
in the most significant motions after the previous cutoff criteria.

%%%%   Figure 6

\begin{figure*}
    \centering
	\includegraphics[width=.5\linewidth, valign=t]{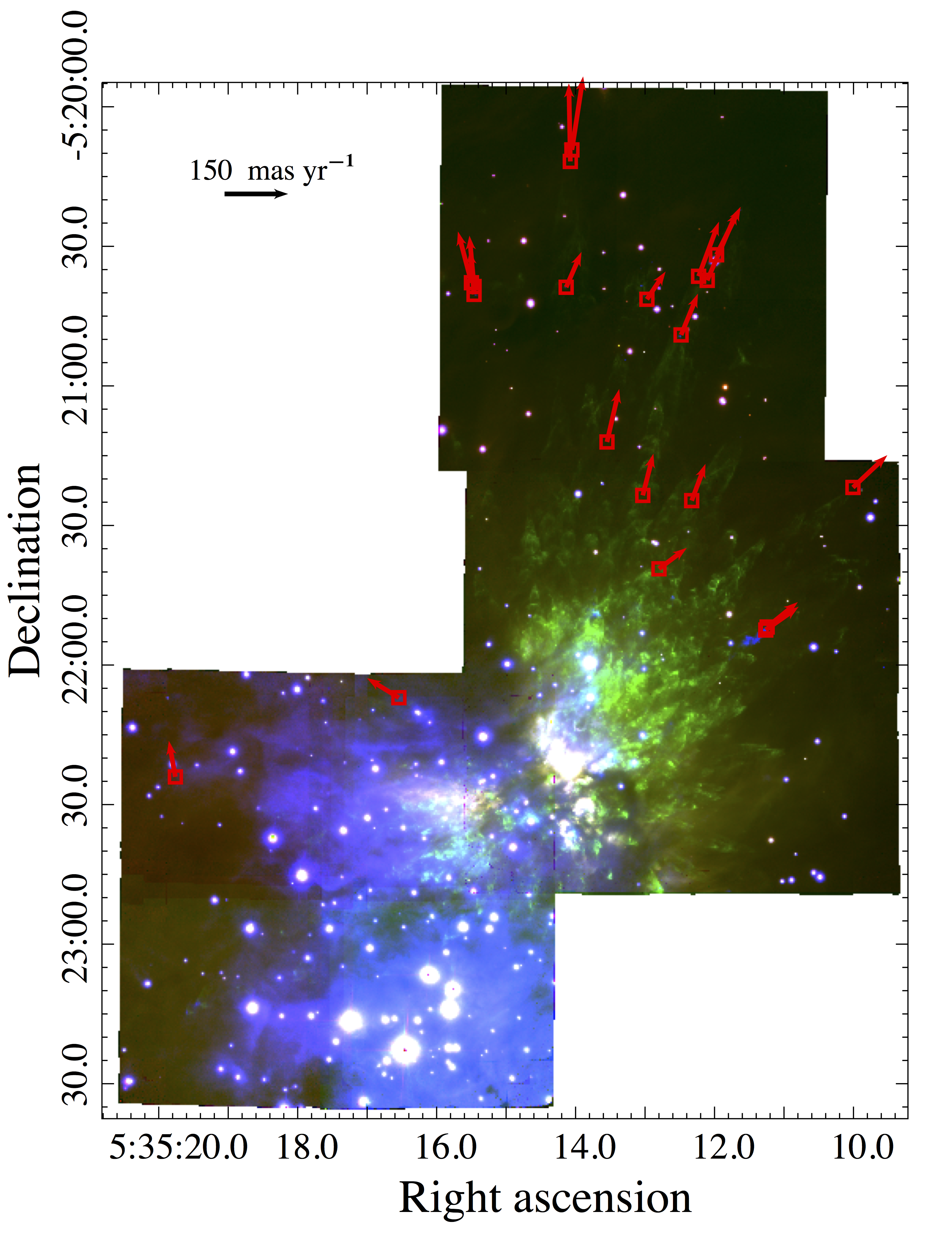}
	\includegraphics[width=.49\linewidth, valign=t]{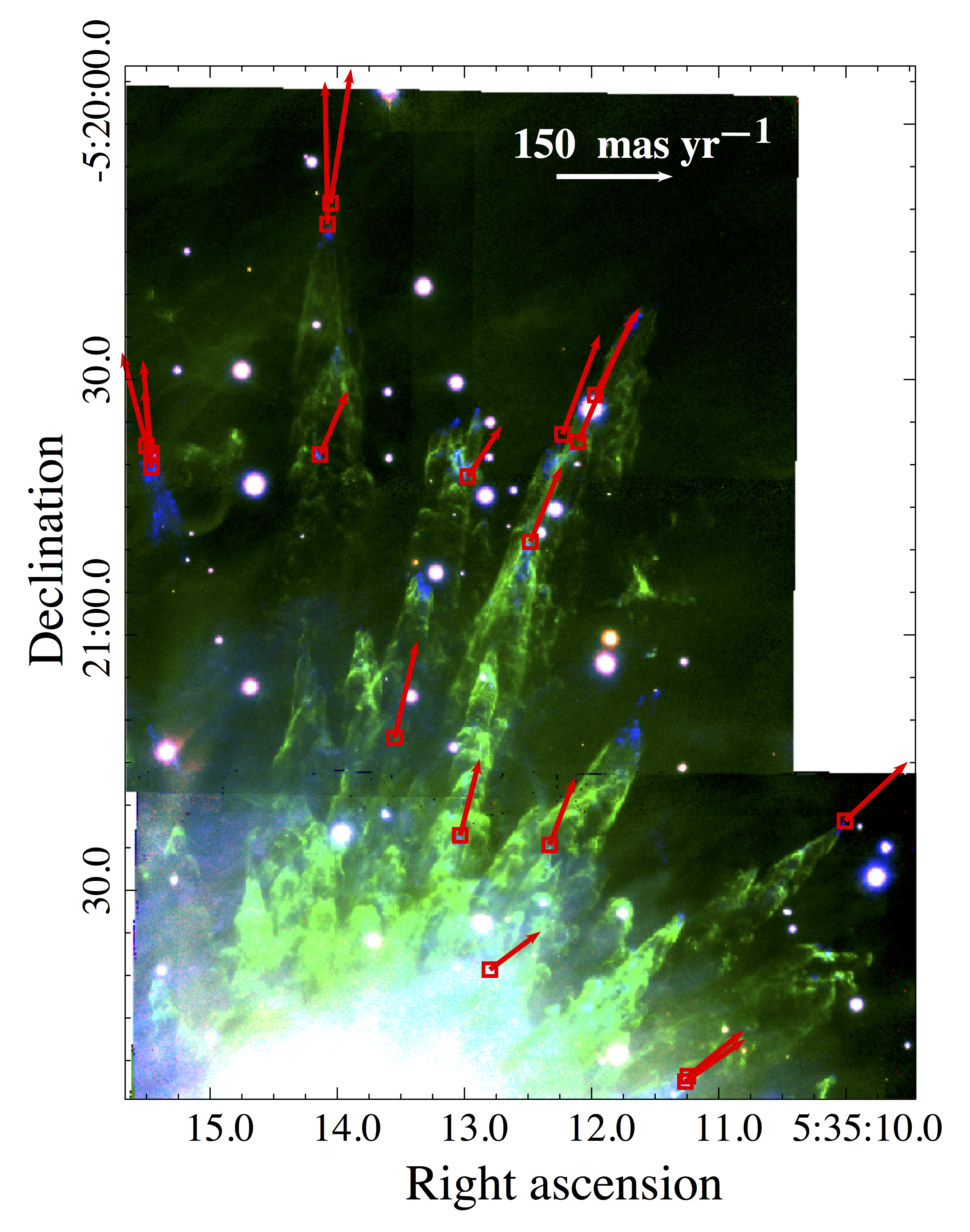}
   \caption{Color composite image of the GSAOI broad-band $K_s$ filter (red), the 2.12 $\mu$m H$_2$ (green) 
   and 1.64 $\mu$m [Fe {\sc ii}] (blue) narrow-band filters towards the OMC1 outflow \citep{bally2015}. (left) Wide-filed image of the OMC1 outflow. (right) Close-up of the most prominent Orion finger systems in the northernmost field in the left panel (same scale, higher contrast). The positions of sources from the high proper motion sample is indicated with red symbols together with their proper motion vectors.}
    \label{bnkl_whole}
 \end{figure*}

The final sample of significant motions is listed in Table \ref{PM_tab}. It consists of 22 sources with transverse velocities in the range of $\sim$95$-$373~km~s$^{-1}$ with most of them lying along the "finger" shaped features towards the north and northwest part of the BN/KL outflow from the Orion OMC1 cloud core \citep{zapata2009, bally2015} and also known as "Orion Fingers" \citep{taylor1984, allen1993}. Figure \ref{bnkl_whole} shows a high-resolution NIR image towards the OMC1 outflow adapted from \citet{bally2015} with the broad-band $K_s$ filter in red, the 2.12 $\mu$m H$_2$ and 1.64 $\mu$m [Fe {\sc ii}] narrow-bands in green and blue, respectively. The OMC1 outflow has an explosive morphology of molecular material consequence of the dynamical encounter of stars which were ejected with speeds of a few tens of km/s more than 500 years ago \citep{bally2005, rodriguez2005}. Most of the material seen in near-IR image in Figure \ref{bnkl_whole} seems to have been triggered by this event. This finding proves the sensitivity of these VLA observations to non-stellar radio emission and the utility of proper motions as an additional piece of information in counterpart identification.

%%%%   Figure 7

\begin{figure*}
    \centering
 	\includegraphics[width=.505\linewidth]{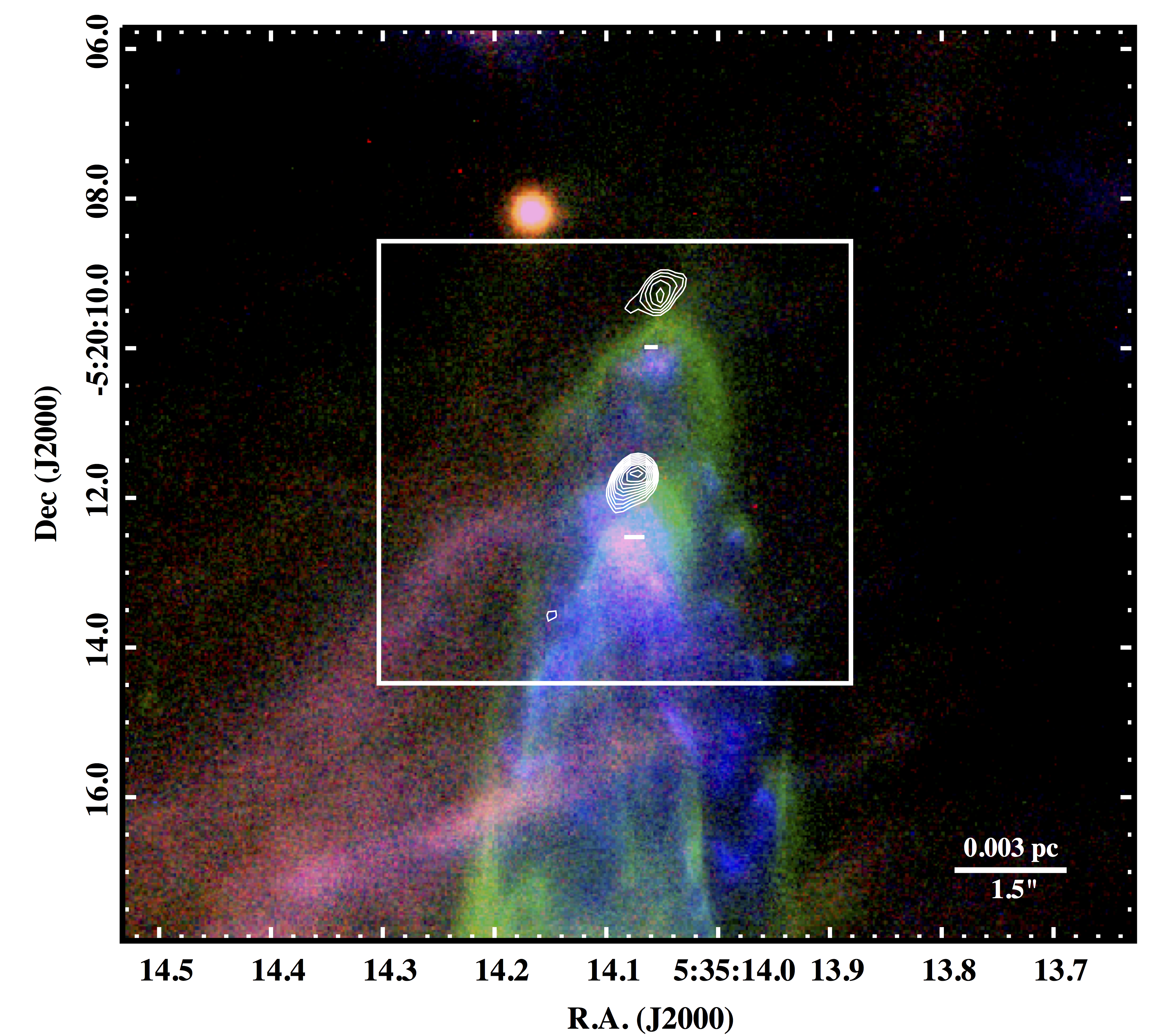}
	\includegraphics[width=.485\linewidth]{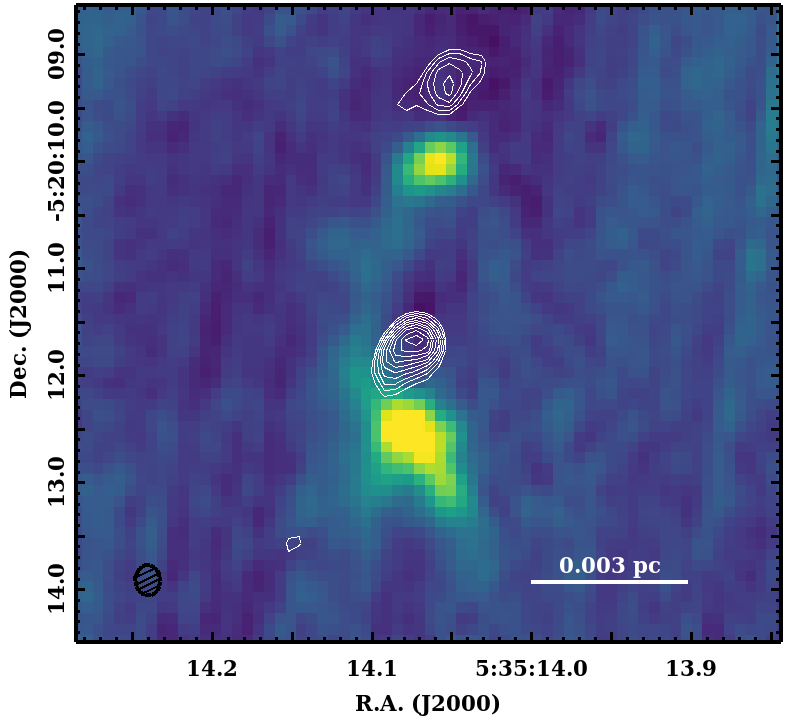}
    \caption{\textit{(left)} Close-up from Figure \ref{bnkl_whole} around the fastest moving sources 153 and 155 (the northernmost in Figure \ref{bnkl_whole}) highlighting the extended emission features in the NIR. The white box indicates the FOV shown in the right panel. \textit{(right)} Background VLA radio continuum map from epoch 2012 centred around the sources in the left panel showing the VLA 2016 epoch contours in white with 10 levels between $65 - 140$ $\mu$Jy~bm$^{-1}$. The synthesized beam size is shown in black in the lower left corner.}
    \label{fastest}
 \end{figure*}

 The sample of fastest proper motions are indicated in yellow symbols in Figure \ref{bnkl_whole}. An example of two of the fastest features detected are shown in Figure \ref{fastest}. These are the sources 153 and 155 and are the northernmost sources seen in Figure \ref{bnkl_whole}, with transverse velocities of $V_T=350 - 373$~km~s$^{-1}$ and uncertainties on the order of 13~km~s$^{-1}$. The left panel of Figure \ref{fastest} shows a close-up of Figure \ref{bnkl_whole} around the sources 153 and 155 in a slightly larger FOV compared to the right panel FOV (white box) to highlight the larger infrared structure of these features. The right panel shows the VLA detections with the 2016 epoch in white contours of 10 levels between $65 - 140$ $\mu$Jy~bm$^{-1}$ and the background image is the epoch 2012. Also, in the left panel the extrapolated position of both radio sources to the time of the NIR observations are indicated in small white symbols (correspond to the propagated position errors marginally seen in Declination and only visible in RA). These are indicated to elucidate where the radio emission comes from. These features appear to coincide with Fe {\sc ii} emission, perhaps suggestive of free-free radio emission. At this point, however, the emission mechanism remains unclear, not least since the emission is too faint for spectral index analysis. 
 
%%%%   Figure 8

\begin{figure*}
    \centering
	\includegraphics[width=\linewidth]{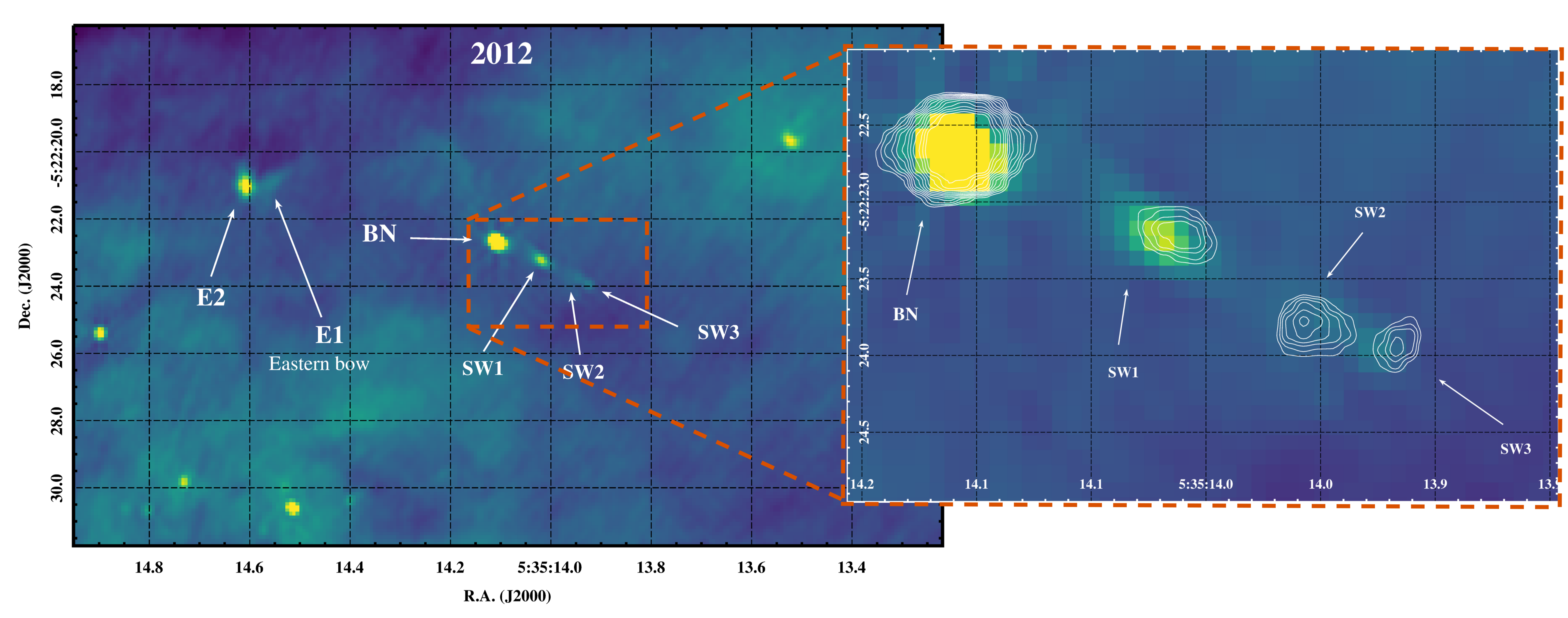}
    \caption{\textit{(left)} VLA radio continuum map from epoch 2012 centred around the source BN. \textit{(right)} Close-up of the BN source and its jet-like features showing the VLA 2016 epoch contours in white with 12 levels between $0.05 - 0.6$~mJy/beam. The proper motion properties of these sources are listed in Table \ref{BN_PM}.}
    \label{bn_zoom}
 \end{figure*}

Additional examples of non-stellar compact radio emission are those associated with the jet-like features around the BN object and shown in Figure \ref{bn_zoom}. The radio continuum map corresponds to the 2012 VLA observations from \cite{for16} while the white contours in the close-up around BN and the southwest features correspond to the 2016 VLA observations. These jet-like features are labeled following the discussion in \cite{bally2020} as E2 and E1 east from BN, and SW1, SW2 and SW3 southwest from BN. The spatial distribution and proper motions of these sources can be interpreted, as further discussed in \cite{bally2020}, to be tracing a low-velocity outflow from BN, although it has been suggested that it may contain a low-luminosity protostar or to be tracing ejected clumps produced by the OMC1 explosion \citep{dzib17, bally2020}. Both sources SW1 and SW3 are moving away from BN towards the west while E2 is moving away from BN towards the east. The absolute PM of these sources are listed in Table \ref{BN_PM} also indicating their equivalent transverse velocities for an adopted distance of 400 pc. The source SW1 (Zapata 11; \citealt{zap04}) was originally reported by \cite{men95} and its PM have been well constrained on longer timescales using up to 9 VLA epochs ranging from 1985 to 2018 \citep{dzib17, rodriguez2020} reporting PM consistent with our measurements. Additionally, \cite{for16} reported spectral index measurement for this source with negative value of $-0.3\pm0.09$ suggesting nonthermal emission, which has been actually found for YSO jets (see Section 6 in \citealt{anglada2018} and reference therein). Source E1 is not included neither in \cite{for16}, since it appears as an extended emission in the epoch 2012, nor in the catalog presented in this work since it is too faint in epoch 2016, however, it is still possible to extract its radio properties in both epochs for the purpose of the discussion in \cite{bally2020}, although with comparatively high positional uncertainties (55-90~mas in $\delta$ and 75-120~mas in $\alpha$). Interestingly, source E1 and SW2 present motions moving back towards BN, however this may be an apparent motion due to intensity variations of different parts in these features, which also leads to a very high PM for E1. The PM measurements for the sources BN, SW1 (Zapata 11) and E2 (IRc23) are consistent with those reported in \cite{rodriguez2020}.

%%%%   Figure 9

%%%%   Table 4
  \begin{table*}
  \centering
  \small
  \caption{Absolute Proper Motions of the BN source and the jet-like features around it and shown in Figure \ref{bn_zoom}.}
  \label{BN_PM}
  \begin{threeparttable}
  \begin{tabular}{cccrrccccc}
  \hline
  \hline
  Source & $\alpha(2000)$  & $\delta(2000)$ &\multicolumn{1}{c}{$\mu_\alpha \cos(\delta)$}&\multicolumn{1}{c}{$\mu_\delta$} & $V_t$ & P.A. & Source ID $^a$ & Comments\\
         & $({\rm \ ^h\ ^m\ ^s})$ & $(\degr \ \arcmin \ \arcsec)$ &\multicolumn{1}{c}{(mas yr$^{-1})$} &\multicolumn{1}{c}{(mas yr$^{-1})$} & (km s$^{-1})$ & $(\degr)$ &  & \\
  \hline
  BN      & 05:35:14.10681$\pm$0.00009 & -5:22:22.6396$\pm$0.0014 & $-6.4\pm0.4$   & $10.1\pm0.4$ & $22.6\pm0.7$ & $328\pm2$ & 156 & \\
  \hline
  SW1     & 05:35:14.01394$\pm$0.00182 & -5:22:23.1999$\pm$0.0188 & $-24.7\pm7.0$   & $ 4.4\pm4.0$ & $47.5\pm12.5$ & $280\pm9$ & 152 & Zapata 11 $^b$\\
  SW2     & 05:35:13.95308$\pm$0.00095 & -5:22:23.8168$\pm$0.0159 & $ 21.0\pm6.2$   & $-7.7\pm4.3$ & $42.3\pm8.1$ & $290\pm8$ & 150 & \\
  SW3     & 05:35:13.91647$\pm$0.00069 & -5:22:23.9290$\pm$0.0155 & $-33.4\pm3.3$   & $11.7\pm3.7$ & $67.1\pm5.1$ & $289\pm5$ & 147 & \\
  \hline
  E1 $^c$  & 05:35:14.53241$\pm$0.00810  & -5:22:20.7142$\pm$0.0864 & $-141.1\pm48.2$ & $96.5\pm34.6$ & $324.2\pm39.5$ & $304\pm8$ & --- & \\
  E2      & 05:35:14.61240$\pm$0.00067 & -5:22:20.9111$\pm$0.0136 & $ 12.6\pm2.7$   & $21.2\pm3.7$ & $46.7\pm6.0$ & $31\pm6$ & 177 & IRc23 $^b$\\
  \hline
  \end{tabular}
  \centering
  \begin{tablenotes}
  \item $^a$ This work (see Table \ref{catalog_tab}).
  \item $^b$ Source designation in \cite{dzib17}.
  \item $^c$ Source properties does not meet the criteria for the catalogs defined by neither \cite{for16} (it's a extended emission) nor in this work (it has SNR$<$5).
  \end{tablenotes}
  \end{threeparttable}
  \end{table*}
  
\subsubsection{Peculiar motion of source 117: COUP 510}\label{coup510}

In the sample of high proper motions there is an intriguing stellar source 
with apparent fast motion of $\mu_\alpha cos(\delta) = -67.0\pm1.7$~mas~yr$^{-1}$ 
and $\mu_\delta = 55.7\pm2.2$~mas~yr$^{-1}$ equivalent to an unusual stellar 
transverse velocity of $V_T=165$~km~s$^{-1}$. It has an X-ray counterpart in 
the COUP survey (COUP 510) presenting high extinction with hydrogen column 
density log$(N_H)=23.54\pm0.06$~cm$^{-2}$ \citep{get05b} and no detection in 
the VISION Ks-band thus presenting characteristics of a deeply embedded object. 
If this corresponds to a genuine linear motion it would be possible to detect 
its expected X-ray position at the time of the COUP observations back in 2003 
by extrapolation, however, in the X-ray COUP images there is no detection at the 
extrapolated position in 1" around apart from the source COUP 510 
and the next closest X-ray source is COUP 533 at $3\farcs1$ away.
Therefore a linear motions is not a plausible scenario for this source's motion. 
We then looked into the VLA archive for high angular resolution multiepoch observations towards this source and found 6 additional observations. These observations have been previously reported by \citet{gomez2005, gomez2008}, and we followed their data calibration and imaging procedures. The phase centers of these observations are consistent with the one used in this work, except for two observations with offsets of 30$''$ (1991.68) and 45$''$ (2004.84), which are still smaller than the primary beam size at the observed frequency. The data calibration follows the standard procedure recommended for the pre-upgrade VLA, and the imaging discarded visibilities provided by baselines longer than 100~k$\lambda$ to diminish the noise caused by the poorly mapped extended emission (for details see \citealt{gomez2008}). The positions were corrected to consider the updated position of the phase calibrator.  In total, the radio source is detected in 8 epochs above 5$\sigma$, with rms noise levels ranging from 3~$-$~80 $\mu$Jy~bm$^{-1}$, spanning almost 25 years including the epochs 2012 and 2016 used in this work.  The main image parameters of these observations are listed in Table \ref{multiepoch_vla_table} together with the measured positions of the source.

%%%%   Table 5
\begin{table*}
\centering
\caption{Multiepoch VLA dataset used for PM measurements of COUP 510.}
\label{multiepoch_vla_table}
\begin{threeparttable}
\begin{tabular}{c c c c c c c}
\hline\hline
        & $\lambda$&    Synthesized Beam size   &  PA     &   RMS   &\multicolumn{2}{c}{Position $^a$}          \\
Epoch   &   (cm)   &  ($B_{maj}\times B_{min}$) &($\degr$)& ($\mu$Jy bm$^{-1}$) & $\alpha(2000)$ & $\delta(2000)$ \\
\hline           
1991.68 &    1.3  &$0\farcs26 \times 0\farcs25$&  $-55$  &  80 & 12$\fs$9497 $\pm$ 0$\fs$0009 &54$\farcs$8044 $\pm$ 0$\farcs$0285   \\
1994.32 &    3.6  &$0\farcs22 \times 0\farcs20$&  $+9$   &  43 & 12$\fs$9797 $\pm$ 0$\fs$0007 &54$\farcs$9168 $\pm$ 0$\farcs$0205    \\
1995.55 &    3.6  &$0\farcs26 \times 0\farcs22$&  $+34$  &  74 & 12$\fs$9648 $\pm$ 0$\fs$0024 &54$\farcs$4929 $\pm$ 0$\farcs$0319    \\
2000.87 &    3.6  &$0\farcs24 \times 0\farcs22$&  $+3$   &  45 & 12$\fs$9817 $\pm$ 0$\fs$0006 &54$\farcs$9118 $\pm$ 0$\farcs$0140  \\
2004.84 &    3.6  &$0\farcs22 \times 0\farcs20$&  $-6$   &  29 & 12$\fs$9814 $\pm$ 0$\fs$0008 &54$\farcs$9668 $\pm$ 0$\farcs$0150   \\
2006.36 &    3.6  &$0\farcs26 \times 0\farcs22$&  $-2$   &  51 & 12$\fs$9825 $\pm$ 0$\fs$0009 &54$\farcs$8534 $\pm$ 0$\farcs$0285   \\
2012.75 &    6    &$0\farcs30 \times 0\farcs19$&  $+30$  &  3  & 12$\fs$9834 $\pm$ 0$\fs$0001 &54$\farcs$9520 $\pm$ 0$\farcs$0029    \\
2016.90 &    6    &$0\farcs40 \times 0\farcs28$&  $-29$  &  33 & 12$\fs$9652 $\pm$ 0$\fs$0004 &54$\farcs$7212 $\pm$ 0$\farcs$0083   \\
\hline
\end{tabular}
\centering
  \begin{tablenotes}
  \item $^a$ $\alpha(2000)=05^h35^m$ ; $\delta(2000)=-05\degr23\arcmin$.
  \end{tablenotes}
\end{threeparttable}
\end{table*}

\begin{figure*}
    \centering
	\includegraphics[width=.49\linewidth]{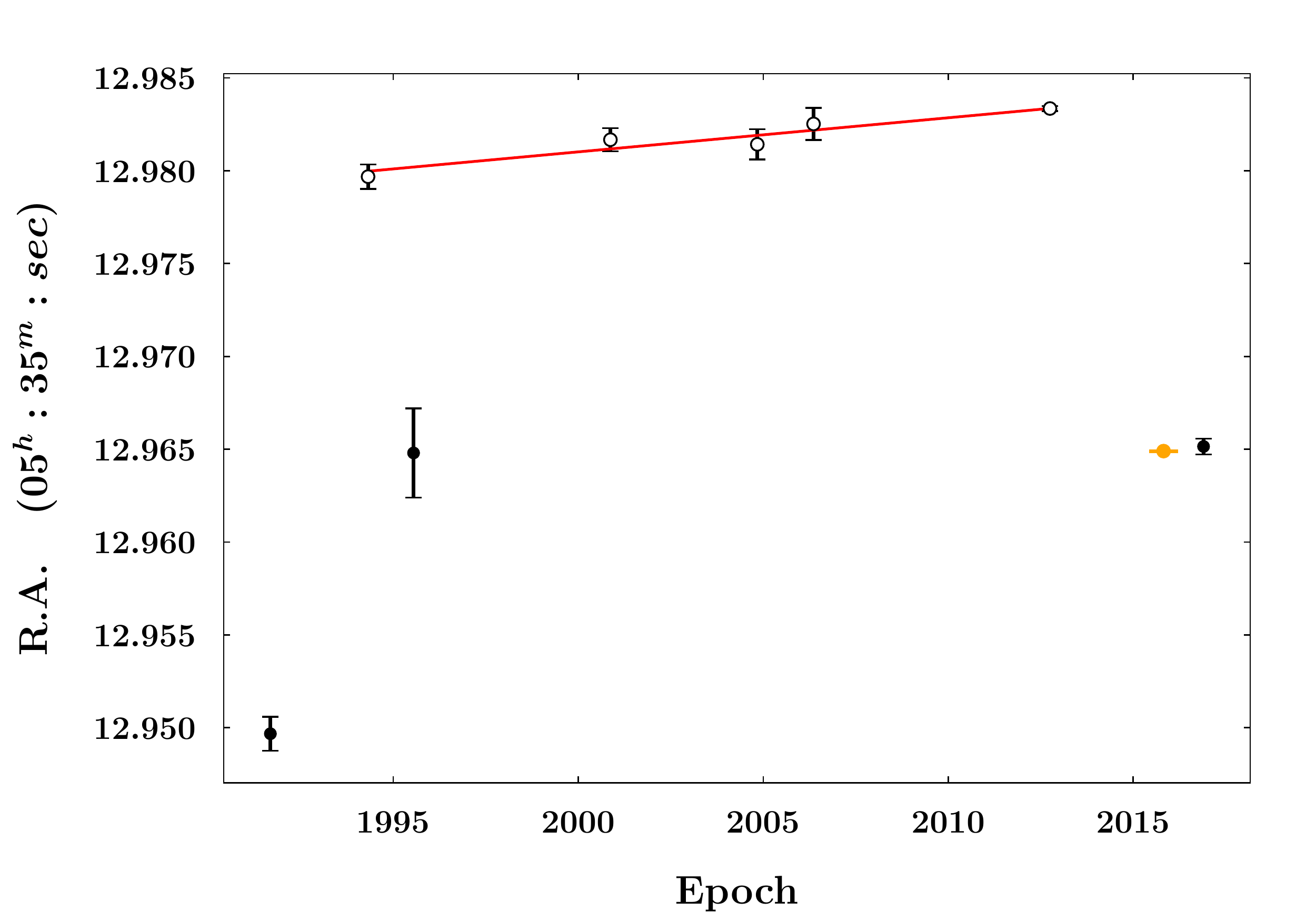}
 	\includegraphics[width=.49\linewidth]{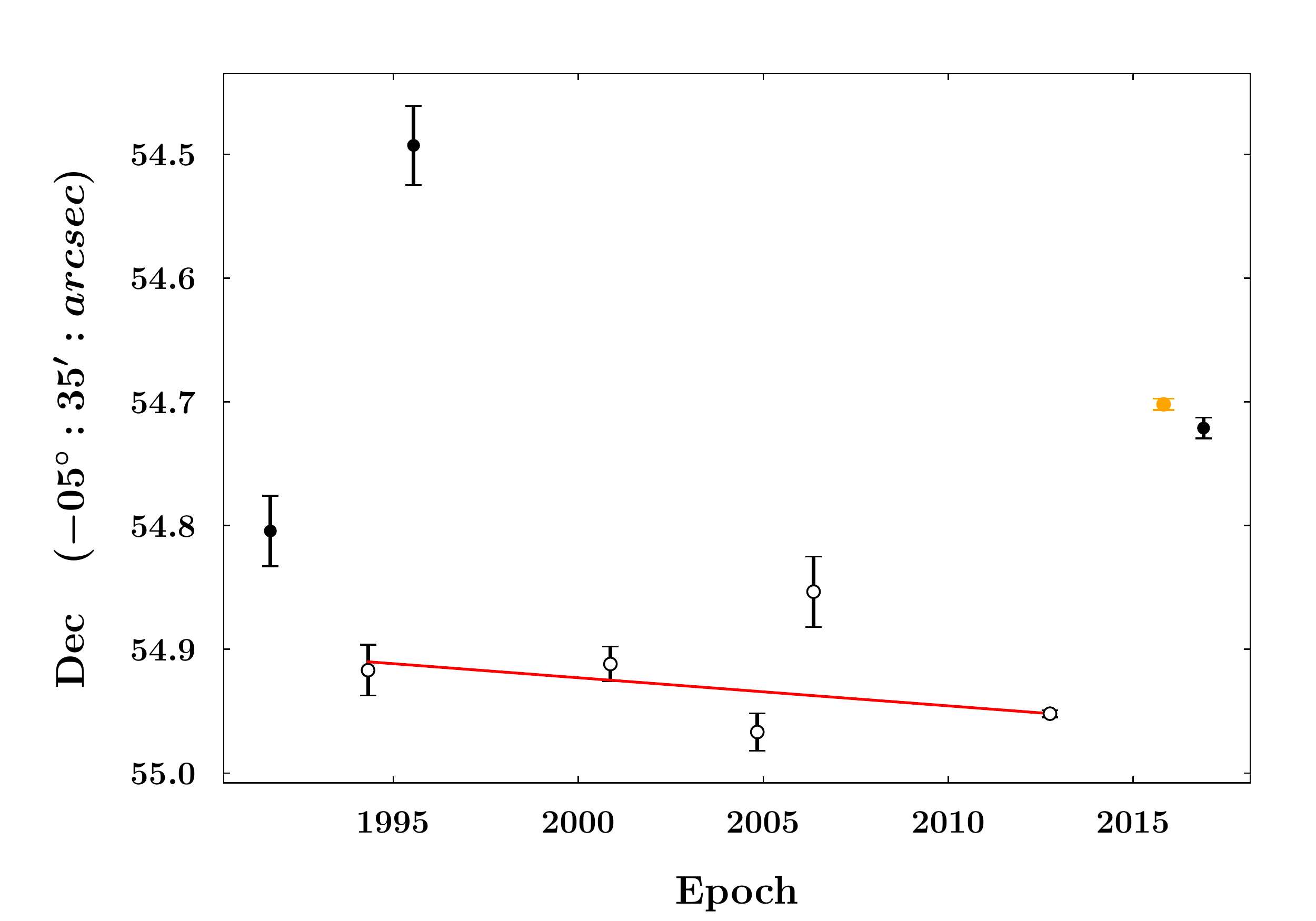}
    \caption{Proper motions in RA and DEC of the source 117 (COUP 510). VLA observations are 
    indicated in black and the VLBA observation is marked in orange. Open black symbols represent 
    one of the putative components of the system used for the hypothetical least-square fit indicated in red.}
    \label{multiepoch_vla}
 \end{figure*}

Figure \ref{multiepoch_vla} shows the PM diagram in RA (left) and DEC (right) of the source 117 using the multiepoch observations listed in Table \ref{multiepoch_vla_table}. We include the additional VLBA observation from 2015 \citep{forbrich2020, dzib2020}. These additional data show that the linear PM estimated only from the two VLA epoch 2012 and 2016 is not representative and may indeed correspond to different components of what seems to be a binary or higher order stellar system. Proper motion measurements for this source have been previously reported in \cite{dzib17} using multi-epoch VLA observations with a time baseline of $\sim$29 years. The reported PM in their work are $\mu_\alpha cos(\delta) = 1.9\pm 1.5$~mas~yr$^{-1}$ and $\mu_\delta = 4.2\pm 5.6$~mas~yr$^{-1}$ which is compatible with a non moving source. The VLBA detection included in this work is compatible with the detection from VLA 2016. What is still intriguing is the fact that in every single observation only one component was detected and never both or more of them at a time, and in four visits with VLBA it was only detected once. Orbital details of this system thus remain unclear.

%%%%   Figure 10

%%%%%%%%%%%%%%%%%%%                RADIO VARIABILITY              %%%%%%%%%%%%%%%%%%%%%%
\subsection{Radio Variability}\label{variability}

The study presented in \cite{for17} using the deep VLA observations showed evidence of variability at very short timescales in the range of minutes with a few cases showing changes in flux density by a factor of 10 in less than 30 minutes. Based on these results we re-imaged the central pointing of our data into time slice images of $\sim5$ minutes integration time following the procedure described in section \S \ref{obs} and then produced light curves for all the 272 sources in the nominal catalog of the central pointing by extracting their flux information from each of the individual 5-minute images using the methodology described in section \S \ref{source_detection}. This time resolution also include exactly 3 science scans to ensure an even time on target throughout the time series. It resulted in a total of 41 individual images with a mean rms noise of $29\,\mu$Jy~bm$^{-1}$ ranging between $26 - 36\,\mu$Jy~bm$^{-1}$. The first important issue here is the increase in the mean rms noise compared with the averaged $\sim$ 5~h image on where the overall rms noise is $10.24\,\mu$Jy~bm$^{-1}$ (see Table \ref{obs_and_images}). Regardless of the decrease in sensitivity all the sources were detected in at least one of the individual images, 46$\%$ of them were detected in at least half of the 41 images and 22$\%$ of them were detected in every single image.

Prior to the quantification of flux variation it was necessary to look at any systematic fluctuation. A very good test-case is the high-mass young stellar object BN, a non-variable thermal radio emitter \citep{for08, for16} with a peak flux density in the averaged 5 hrs image of $S_{\nu} = 2.370\pm0.023$~mJy~bm$^{-1}$. Its peak flux density from the light curves has a mean value of $S_{\nu} = 2.354\pm0.007$~mJy~bm$^{-1}$ with a relative standard deviation of $\sim5\%$ ($0.127$~mJy~bm$^{-1}$). This systematic variation is smaller than the typical value for even just moderate variable sources with relative standard deviation $>$20\% and therefore it does not impact our results of variability analysis. This measure for the BN source is also consistent with its peak flux density observed in the deep VLA catalog of $S_{\nu} = 2.3413\pm0.0026$~mJy~bm$^{-1}$ which represent a variation of just $0.9\pm0.7\%$.

%%%%   Figure 11
\begin{figure*}
    \centering
	\includegraphics[width=0.45\linewidth]{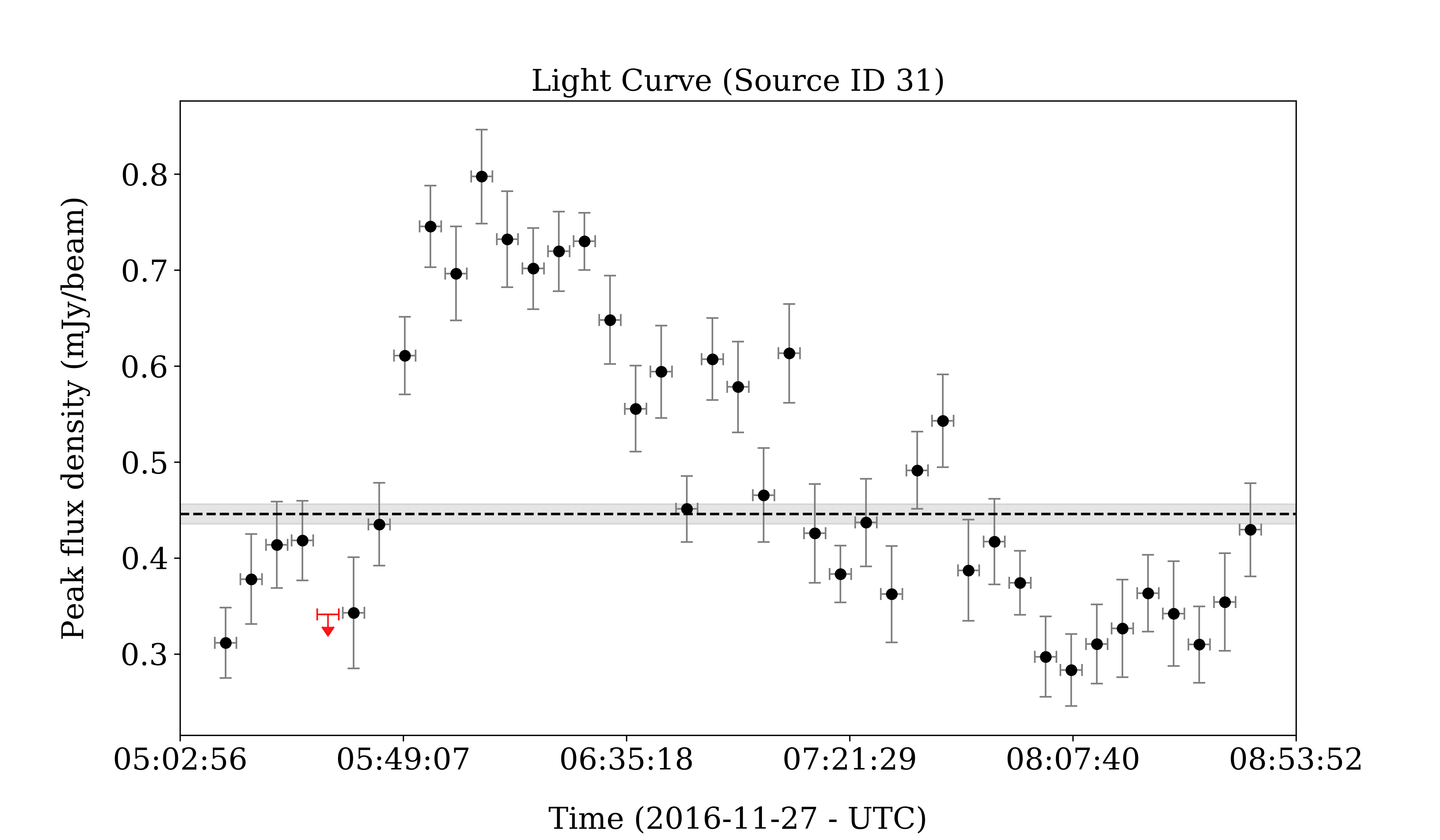}
	\includegraphics[width=0.45\linewidth]{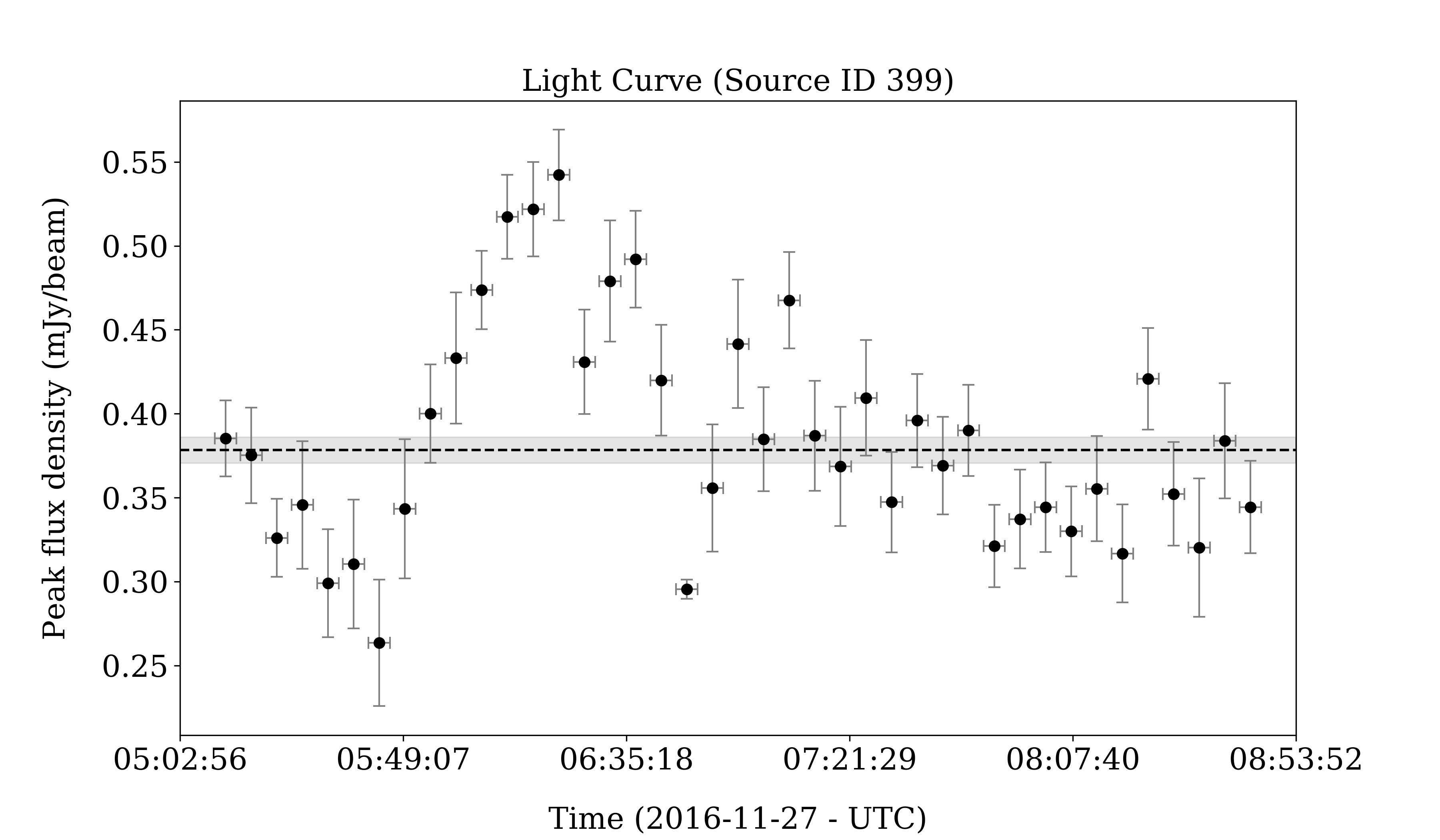}
	\includegraphics[width=0.45\linewidth]{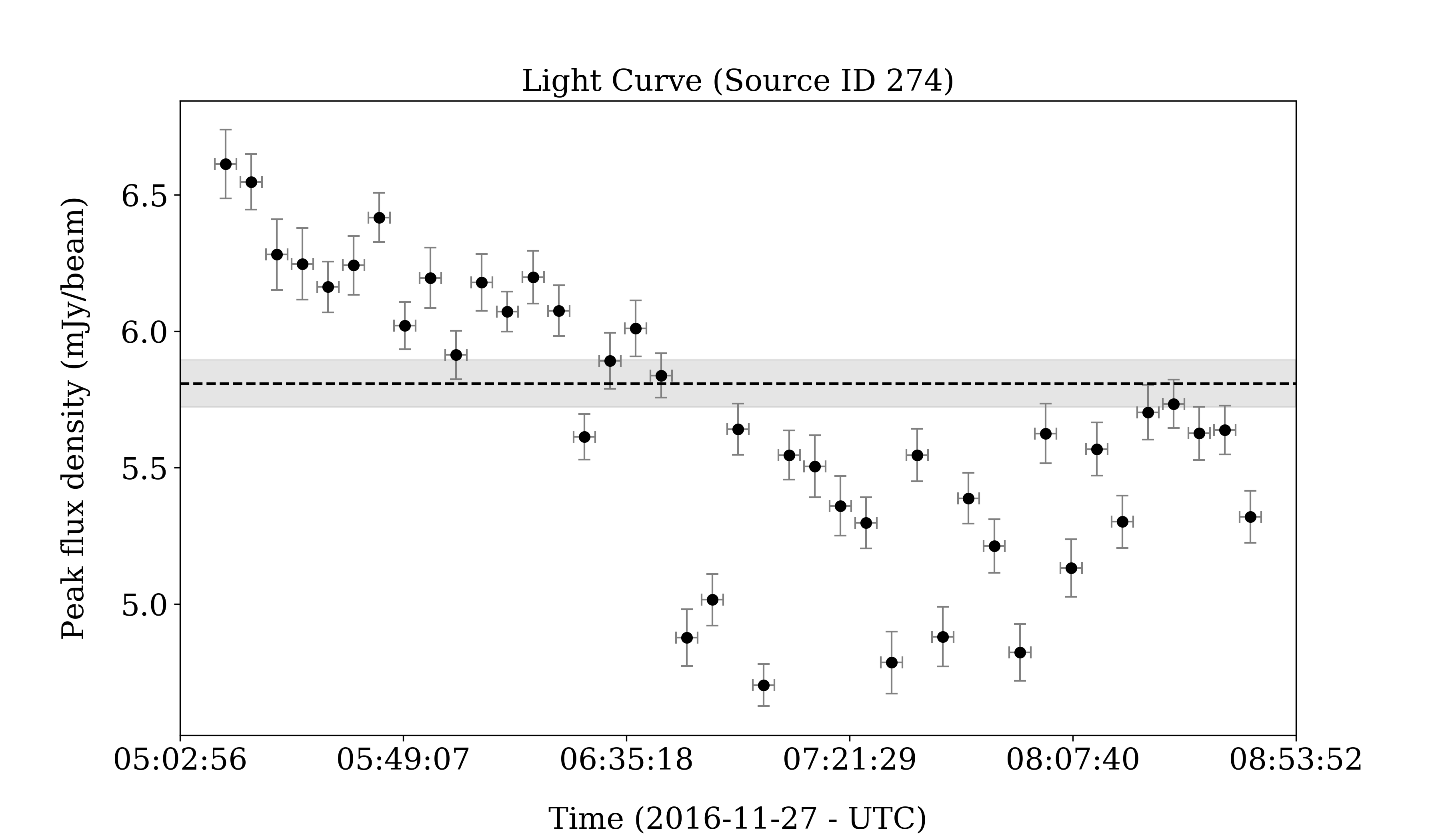}
	\includegraphics[width=0.45\linewidth]{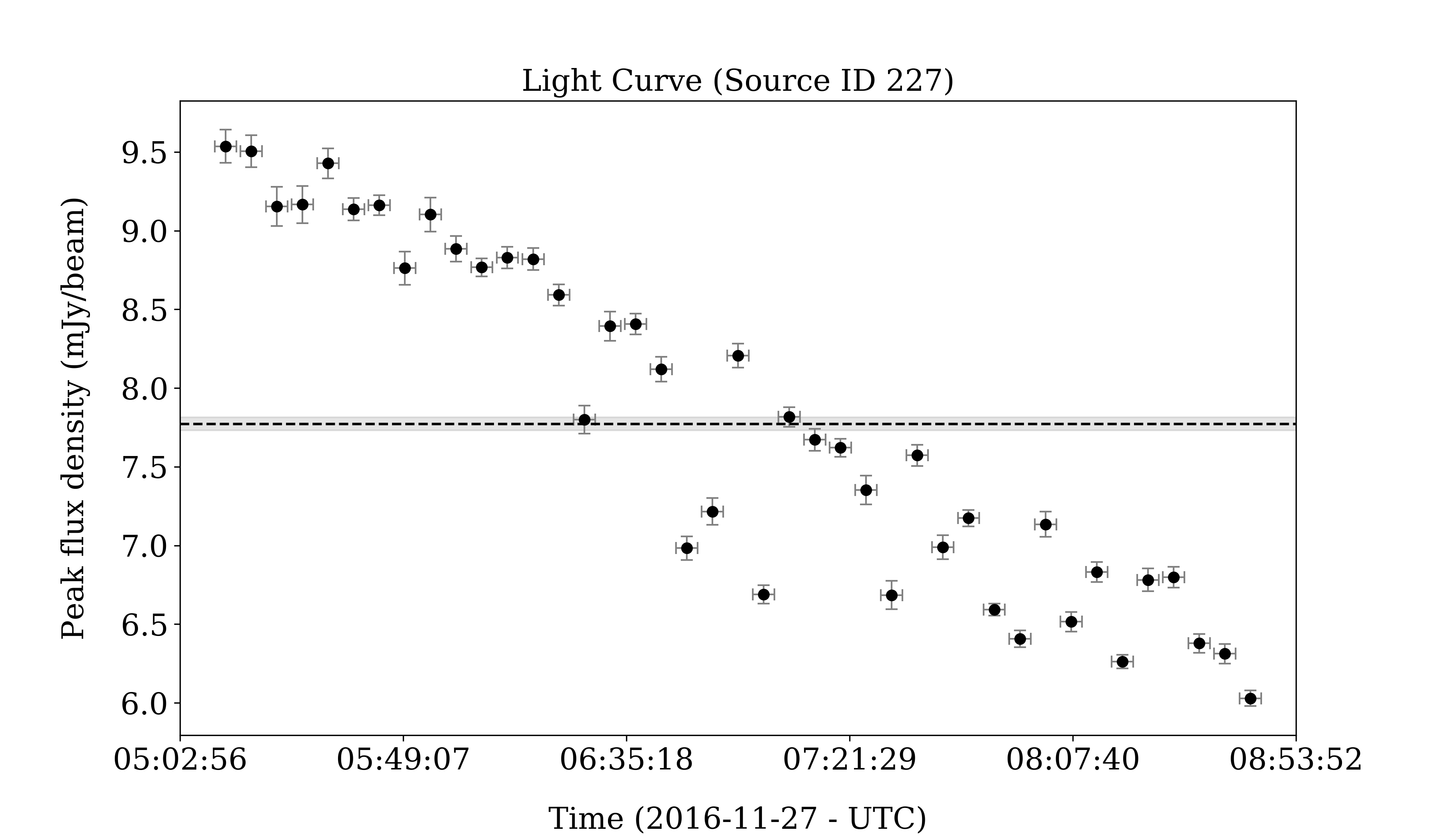}
	\includegraphics[width=0.45\linewidth]{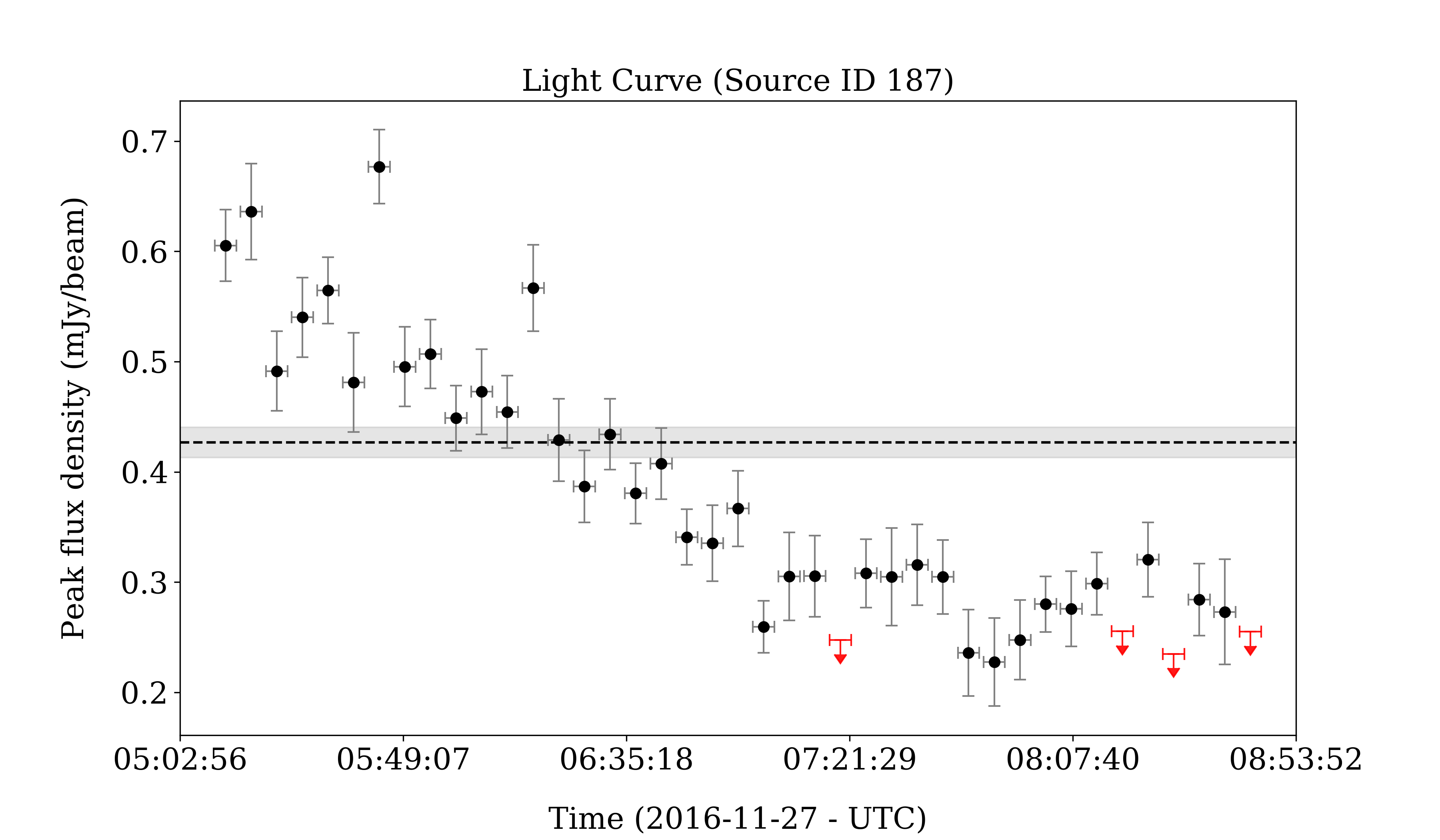}
	\includegraphics[width=0.45\linewidth]{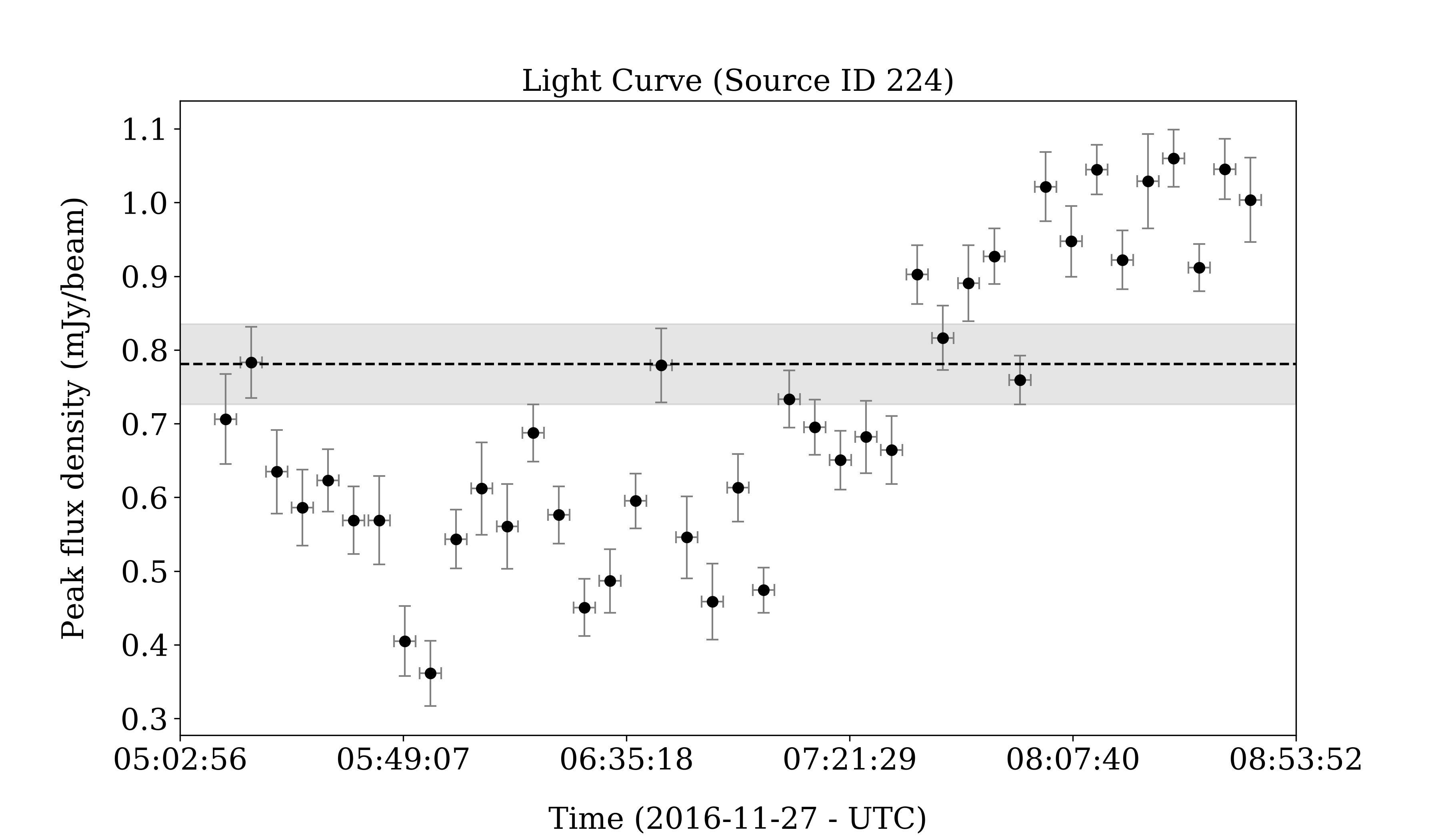}
    \caption{Radio light curves at 5 minutes time resolution. Detections are indicated in black symbols together with its uncertainties. Red symbols indicate the 5$\sigma$ upper limits when the source is not detected. The peak flux density from the averaged 5 hrs image is shown by the horizontal dashed black line together with its uncertainty in grey horizontal band. All the measurements are corrected by the primary beam response.}
    \label{LC}
\end{figure*}

A few examples of light curves are shown in Figure \ref{LC} for sources presenting a small flare-like event and others presenting either increasing or decreasing light curves. The peak flux density of a source in the averaged long exposure image evidently does not always represent the overall evolution of its radio emission and there is a clear fluctuation at short time scales of just minutes that even not being extreme events can still be considerably brighter than the averaged peak flux density. Following \citet{for17}, an extreme variability event refers to a change in flux density by a factor of at least 10 in less than an hour. We define a variability factor VF as the ratio between the maximum and the minimum peak flux density in the timeseries dataset. Although, there is evidence of variability in several sources in the sample, none of them show changes in flux density above a factor of 5 and indeed $\sim$85\% have VF$<$3.

The occurrence rate of extreme radio variability in the ONC has been estimated by \cite{for17} based on the deep observations where they found a mean time between these events of 2220$\pm$1280~h ($\sim3$ months). The sources involved in this estimate are only sources with X-ray counterpart in the COUP survey assuming that this is the most complete sample of YSOs in the inner ONC. In our sample only 157 out of 272 sources have X-ray counterparts in the COUP survey, and the cumulative radio observing time of these sources is 785 h (5 h on each of the 157 sources) for which we found no extreme variability (not by a factor $\geq$10). However, this result lies within the probabilities of only one such event in a cumulative radio observing time of 2220$\pm$1280~h and yet, our cumulative radio observing time is still below the lower limit of 940 h considering the error. The new cumulative radio observing time adds up to 7445 h including the deep \citep{for16} and the new observation (only the central pointing), it leads to a mean time between extreme radio variability events of 2482$\pm$1433~h. This larger uncertainty compared to the previous estimate is a result of the same number of events we are still considering (3 extreme events from \citealt{for17}).

%%%%   Table 6
\begin{table*}
\centering
\caption{Main parameters for the sources with highest VF between the two VLA epochs.}
\label{extreme_VF}
\begin{threeparttable}
\begin{tabular}{c c c c c c c c}
\hline\hline
Source ID$^a$ &[FRM2016]$^b$& VF & $S_{\nu}$ $^a$ & $S_{\nu}$  $^b$ &Sp. index $^b$ & COUP & log$(N_H)$ $^c$\\
          &     &                &\multicolumn{2}{c}{(mJy bm$^{-1}$)}&                &        &   (cm$^{-2}$)    \\
\hline           
  31  &   4 &12.05 $\pm$ 1.76 & 0.446 $\pm$ 0.024 & 0.037 $\pm$ 0.005 & $...$          & 141 & 21.07 $\pm$ 0.03\\
  315 & 378 &10.93 $\pm$ 0.56 & 6.176 $\pm$ 0.312 & 0.565 $\pm$ 0.003 & 0.09$\pm$0.01  & 932 & 21.17 $\pm$ 0.09\\
  179 & 196 & 9.40 $\pm$ 0.64 & 0.715 $\pm$ 0.039 & 0.076 $\pm$ 0.003 & $-0.20\pm$0.23 & 648 & 21.52 $\pm$ 2.26\\
\hline
\end{tabular}
\centering
\begin{tablenotes}
\item $^a$ This work.
\item $^b$ Identification number [FRM2016], flux density and spectral index from \cite{for16}.
\item $^c$ Hydrogen column density from \cite{get05b}.
\end{tablenotes}
\end{threeparttable}
\end{table*}

On the other hand, it is also possible to look for variability on longer timescales by comparing our central pointing against the deep VLA observations. This is a particularly good comparison since these are identical observations and all the sources are affected by the same primary beam response correction due to their equivalent distance to the phase center. In our sample 253 sources are detected in the two datasets. We can measure the variability factor between these two epochs for each source by taking the ratio between their nominal peak flux density from both catalogs. There are two sources presenting a VF$>$10 at this long timescale and a third one marginally below this limit with a VF of 9.4$\pm$0.6. These three sources have X-ray counterparts in COUP with hydrogen column densities in the range log$(N_H)=21.07 - 21.52$ cm$^{-2}$ \citep{get05b}. These parameters are summarized in Table \ref{extreme_VF}.

%%%%   Table 7
\begin{table}
\centering
\caption{Main parameters for the brightest sources in the deep catalog not detected in the new VLA data.}
\label{bright_not_detected}
\begin{threeparttable}
\tabcolsep=0.11cm
\begin{tabular}{c c c c c c}
\hline\hline
 [FRM2016]  &       VF       & $S_{\nu}$ & Sp. index  & COUP   & log$(N_H)$  \\
     &                &(mJy bm$^{-1}$) &                 &        &   (cm$^{-2}$)    \\
(1)&(2)&(3)&(4)&(5)&(6)\\
\hline           
 422 & 17.85 $\pm$ 0.03 & 0.676 $\pm$ 0.003 & 1.1 $\pm$ 0.1  & 997    & 21.60 $\pm$ 0.02   \\
 515 & 28.89 $\pm$ 0.00 & 1.188 $\pm$ 0.004 & 1.4 $\pm$ 0.1  & 1232   & 20.00 $\pm$ 0.00   \\
\hline
\end{tabular}
\centering
\begin{tablenotes}
    \item Columns (1), (3), and (4) are identification number, peak flux density, and spectral index, respectively, from \citet{for16}.
    \item Column (2): VF between the two epochs considering 5 times the local rms in the new observations.
    \item Column (5) and (6) are identification number and Hydrogen column density from \cite{get05b}.
\end{tablenotes}
\end{threeparttable}
\end{table}

Non-detections in the new observations of previously detected sources are not just due to a difference in sensitivity but potentially also due to variability at least for the brightest sources in the deep observations that should be clearly detected in the new VLA data. There are in total 303 sources in the deep catalog not detected in the central pointing of the new VLA data. We looked for their positions in the new data and considered 5 times the local rms within a box of 14 pixels ($1\farcs4$) as upper limits to compare against the peak flux density in the deep catalog. This allows us to estimate a lower limit for their variability factor. As expected, most of the sources have a VF below 10 and are likely too faint for the new observations, but the are two interesting sources with VF of 18 and 29 listed in Table \ref{bright_not_detected}. These two sources are indeed in the list of only 13 extremely variable sources in \cite{for17}. Source [FRM2016]~422 in \cite{for17} shows an extreme radio flare on timescales of $\sim$40~h  while source [FRM2016] 515 shows the most extreme variability within just 30 minutes (by a factor of $\sim$100), also coinciding with an almost simultaneous X-ray flare of similar duration. While source [FRM2016] 422 was also not detected in any of the outer pointings, source [FRM2016]~515 was detected in pointings 3 and 6 with peak flux densities of 0.15$\pm$0.01~mJy~bm$^{-1}$ and 0.18$\pm$0.01~mJy~bm$^{-1}$, respectively, leading to a long term variability with VF$\sim$7 respect to \citet{for17}, and VF$\sim$5 respect to the central pointing (considering the 5$\sigma$ upper limit) which were observed with one month difference. Both sources [FRM2016]~422 and [FRM2016]~515 have reported spectral types K8 and O9.5-B2, respectively \citep{hil13}, although for the high-mass star it is noted that the radio emission may be detected from unresolved lower-mass companion.

%%%%   Table 8
\begin{table}
\centering
\caption{Main parameters for the brightest new detections in the VLA 2016 epoch (S/N$\geq$50).}
\label{new_detections}
\begin{threeparttable}
\begin{tabular}{c c c c c}
\hline\hline
 Source ID &       VF $^a$   & $S_{\nu}$       & COUP   & log$(N_H)$ $^b$  \\
           &                 &(mJy bm$^{-1}$)  &        &   (cm$^{-2}$)    \\
\hline           
   47      & 2.2 & 0.31 $\pm$ 0.02 &  260   & 20.93 $\pm$ 0.04   \\
  438      & 1.5 & 0.69 $\pm$ 0.04 & 1259   & 21.55 $\pm$ 0.05   \\
\hline
\end{tabular}
\centering
\begin{tablenotes}
\item $^a$ From the light curves analysis in this work.
\item $^b$ Hydrogen column density from \cite{get05b}.
\end{tablenotes}
\end{threeparttable}
\end{table}

Finally, another interesting finding concerns the new detections in the central pointing not detected in the previous deep VLA data. There are 19 new detections, 10 of which have X-ray counterparts in the COUP survey. There are 9 sources with S/N$>$10 and two of them have high S/N of 50 and 82. The main parameters of these two bright new sources are listed in Table \ref{new_detections}.

%%%%%%%%%%%%%%%%%%%%%%%%%%%%%%%%%%%%%%%%%%%%%%%%%%%%%%%%%%%%%%%%%%%%%%%%%%%
%%%%%%%%%%%%%%%%%       Summary and CONCLUSIONS        %%%%%%%%%%%%%%%%%%%%
%%%%%%%%%%%%%%%%%%%%%%%%%%%%%%%%%%%%%%%%%%%%%%%%%%%%%%%%%%%%%%%%%%%%%%%%%%%
\section{Summary and Conclusions}\label{conclusions}

We have presented a new deep, high-resolution catalog of compact radio sources towards the ONC at centimeter wavelengths using the most extended configuration of the VLA. This is the deepest catalog for the surrounding areas of the ONC reported to date, reaching rms noise levels between 3$-$5 $\mu$Jy bm$^{-1}$ at distances of $\sim$10$\farcm4$ from the center of the cluster, significantly improving the general census of known compact radio sources in the cluster. We detected a total of 521 radio sources above 5$\sigma$ threshold over an area of $\sim$20$\arcmin\times20\arcmin$. In this catalog, 198 sources are new detections not previously reported at these frequencies. The highest stellar surface density occurs in the inner region, and yet the number of radio sources may still be underestimated due to the difficulty of disentangling small-scale structure of the nebula and stellar point sources.

With our new catalog, we are sensitive not only to stellar radio emission but also to radio emission originating elsewhere in the ONC, for example in outflows and shocks. It turned out in this regard that even the relatively short time baseline of 4.19 years is sufficient to trace high-velocity proper motions in the ONC, which can be more easily measured with phase-referenced radio than with optical observations. We identify radio sources in the deep 2012 catalog as co-moving with and thus originating in ejecta of the OMC1 explosion. While we also appear to detect fast proper motions towards stellar sources, those are likely due to insufficiently sampled multiple systems. 

We find that the central pointing, also covered in the previous deep survey, contains the majority of ONC radio sources, and interestingly, the surface density of radio sources falls off faster than that of X-ray sources, tracing young stars. This may be due to a surplus of thermal/free-free sources in the inner ONC that are ionized by $\theta^1$ Ori C. Additionally, we find that most ($>50$\%) radio sources have X-ray counterparts, throughout the area studied here, while only a minority ($<20$\%) of X-ray sources have radio counterparts. X-ray emission thus remains a poor predictor of radio emission in this sample of young stars. Given the regional differences that we find, this is not simply a sensitivity effect. 

Finally, a radio variability analysis for sources in the inner ONC is presented as a follow-up of our previous deep VLA variability study. We produced radio light curves at high time resolution for sources in the central pointing finding changes in flux density by a factor $\leq$5 on timscales of minutes to a few hours thus we do not find extreme variability events. The majority of the sources have detections in the previous deep VLA observations enabling the study of long term variability where we only find two sources with changes in flux density $\geq$10. Based in these two studies we find a mean time between extreme radio variability events of 2482$\pm$ 1433 h.

%%%%%%%%%%%%%%%%%%%%%%%%%%%%%%%%%%%%%%%%%%%%%%%%%%%%%%%%%%%%%%%%
%%%%%%%%%%%%%%%%%       ACKNOWLEDGEMENTS      %%%%%%%%%%%%%%%%%%
%%%%%%%%%%%%%%%%%%%%%%%%%%%%%%%%%%%%%%%%%%%%%%%%%%%%%%%%%%%%%%%%
\section*{Acknowledgements}

We thank Mark Reid and Karl Menten for helpful discussions and the anonymous referee for a constructive review of the manuscript. This research made use of Astropy,\footnote{\url{http://www.astropy.org}} a community-developed core Python package for Astronomy \citep{astropy13, astropy18} and Matplotlib \citep{hun07}; APLpy, an open-source plotting package for Python (\url{https://github.com/aplpy/aplpy}); The University of Hertfordshire high-performance computing facility (\url{http://stri-cluster.herts.ac.uk}). 

\section{Data Availability}

The VLA data underlying this article are available under the project code 16B-268 from the NRAO archive (\url{archive.nrao.edu}).

%%%%%%%%%%%%%%%%%%%%%%%%%%%%%%%%%%%%%%%%%%%%%%%%%%%%%%%%%%%%%%%%
%%%%%%%%%%%%%%%%%%        REFERENCES        %%%%%%%%%%%%%%%%%%%%
%%%%%%%%%%%%%%%%%%%%%%%%%%%%%%%%%%%%%%%%%%%%%%%%%%%%%%%%%%%%%%%%
\bibliographystyle{mnras}
\bibliography{orion} 

\appendix

\section{List of sources with high proper motions}\label{PM_table_app}

The sample of sources with high proper motions discussed in Section \S\ref{PM_section} is listed in Table \ref{PM_tab} and the sky-projected distribution is shown in Figure \ref{bnkl_whole}. Column (1) shows the identification number from the catalog reported in Table \ref{catalog_tab}. Columns (2) and (3) indicate the measured proper motions in $\alpha$ and $\delta$, respectively. Column (4) indicates the total proper motion in the plane of the sky, column (5) the transverse velocity for an adopted distance of 400~pc to the ONC, column (6) the position angle respect to the north celestial pole (NCP), and column (7) indicates the identification number from \citet{for16}.

%%%%   Table A1
\begin{table*}
\centering
\caption{Sample of high proper motion in the OMC1 cloud core.}
\label{PM_tab}
\begin{threeparttable}
\begin{tabular}{lccccccr}
\hline \hline
ID & $\mu_\alpha \cos\delta$ & $\mu_\delta$ & $\mu_{total}$ &$V_t$  $^a$& PA & [FRM2016]  $^b$ \\
 & (mas yr$^{-1}$) & (mas yr$^{-1}$)&(mas yr$^{-1}$) &(km s$^{-1}$) &($\degr$) & \\
 (1) & (2)  &  (3)  & (4)  & (5) & (6)  &  (7)\\
\hline
63  & -70.7 $\pm$ 4.1  &  66.5 $\pm$ 7.6  &  97.1 $\pm$ 4.5  & 184.1 $\pm$ 8.6  & -46.8 $\pm$ 2.8 & 25 \\
78  & -56.5 $\pm$ 4.1  &  46.0 $\pm$ 4.0  &  72.8 $\pm$ 3.2  & 138.1 $\pm$ 6.0  & -50.9 $\pm$ 2.5 & 46 \\
80  & -61.2 $\pm$ 3.2  &  43.3 $\pm$ 3.4  &  75.0 $\pm$ 2.7  & 142.2 $\pm$ 5.1  & -54.7 $\pm$ 2.1 & 48 \\
91  & -47.1 $\pm$ 8.9  & -22.8 $\pm$ 14.0 &  52.3 $\pm$ 7.6  &  99.2 $\pm$ 14.4 &  64.2 $\pm$ 10.9 & 67 \\
94  & -53.1 $\pm$ 4.7  & 105.8 $\pm$ 6.2  & 118.4 $\pm$ 4.9  & 224.5 $\pm$ 9.3  & -26.7 $\pm$ 2.0 & 70 \\
98  & -75.2 $\pm$ 13.6 & 174.3 $\pm$ 11.0 & 189.8 $\pm$ 8.3  & 359.9 $\pm$ 15.8 & -23.4 $\pm$ 2.9 & 74 \\
100 & -46.6 $\pm$ 4.3  & 124.8 $\pm$ 5.2  & 133.2 $\pm$ 3.63 & 252.6 $\pm$ 6.9  & -20.5 $\pm$ 1.3 & 76 \\
101 & -25.2 $\pm$ 3.9  &  67.3 $\pm$ 4.0  &  71.9 $\pm$ 3.0  & 136.3 $\pm$ 5.6  & -20.5 $\pm$ 2.3 & 78 \\
103 & -33.8 $\pm$ 7.0  &  82.6 $\pm$ 5.9  &  89.2 $\pm$ 4.9  & 169.2 $\pm$ 9.2  & -22.3 $\pm$ 3.5 & 81 \\
111 & -46.0 $\pm$ 7.3  &  34.7 $\pm$ 4.8  &  57.6 $\pm$ 5.5  & 109.3 $\pm$ 10.4 & -53.0 $\pm$ 4.8 & 91 \\
119 & -29.7 $\pm$ 8.2  &  44.1 $\pm$ 8.1  &  53.1 $\pm$ 6.4  & 100.7 $\pm$ 12.2 & -33.9 $\pm$ 7.1 & 97 \\
124 & -20.8 $\pm$ 7.1  &  81.6 $\pm$ 4.5  &  84.3 $\pm$ 3.4  & 159.7 $\pm$ 6.5  & -14.3 $\pm$ 3.6 & 103 \\
136 & -27.1 $\pm$ 4.3  & 117.4 $\pm$ 3.9  & 120.5 $\pm$ 3.3  & 228.5 $\pm$ 6.3  & -13.0 $\pm$ 1.7 & 129 \\
153 & -27.9 $\pm$ 9.8  & 182.4 $\pm$ 8.1  & 184.5 $\pm$ 7.0  & 349.8 $\pm$ 13.3 &  -8.7 $\pm$ 2.6 & 159 \\
155 &   3.7 $\pm$ 6.8  & 196.8 $\pm$ 7.5  & 196.8 $\pm$ 5.7  & 373.2 $\pm$ 10.8 &   1.1 $\pm$ 1.5 & 161 \\
157 & -27.6 $\pm$ 7.9  &  63.6 $\pm$ 9.9  &  69.4 $\pm$ 7.7  & 131.5 $\pm$ 14.6 & -23.5 $\pm$ 5.4 & 164 \\
207 &  10.0 $\pm$ 4.1  & 109.6 $\pm$ 4.3  & 110.1 $\pm$ 3.8  & 208.7 $\pm$ 7.2  &   5.2 $\pm$ 1.9 & 237 \\
208 &   7.8 $\pm$ 6.4  &  88.7 $\pm$ 7.4  &  89.0 $\pm$ 6.5  & 168.7 $\pm$ 12.3 &   5.4 $\pm$ 3.6 & 238 \\
212 &  29.6 $\pm$ 19.0 & 112.4 $\pm$ 6.6  & 116.2 $\pm$ 7.3  & 220.4 $\pm$ 13.9 &  14.8 $\pm$ 8.7 & 239 \\
263 &  56.2 $\pm$ 4.5  &  34.5 $\pm$ 4.3  &  65.9 $\pm$ 3.6  & 125.0 $\pm$ 6.7  &  58.5 $\pm$ 3.0 & 308 \\
372 & -11.3 $\pm$ 4.0  & -62.8 $\pm$ 5.6  &  63.8 $\pm$ 4.4  & 121.0 $\pm$ 8.3  &  10.2 $\pm$ 2.8 & 450 \\
411 &  37.2 $\pm$ 4.8  & -34.1 $\pm$ 6.0  &  50.5 $\pm$ 5.1  &  95.6 $\pm$ 9.7  & -47.5 $\pm$ 5.9 & 495 \\
\hline
\end{tabular}
\centering
\begin{tablenotes}
\item $^a$ For an adopted distance of 400 pc to the ONC.
\item $^b$ \cite{for16}.
\end{tablenotes}
\end{threeparttable}
\end{table*}

\section{KS test on multi-wavelength population analysis}\label{KS_test}

\begin{figure}
    \centering
    \includegraphics[width=\linewidth]{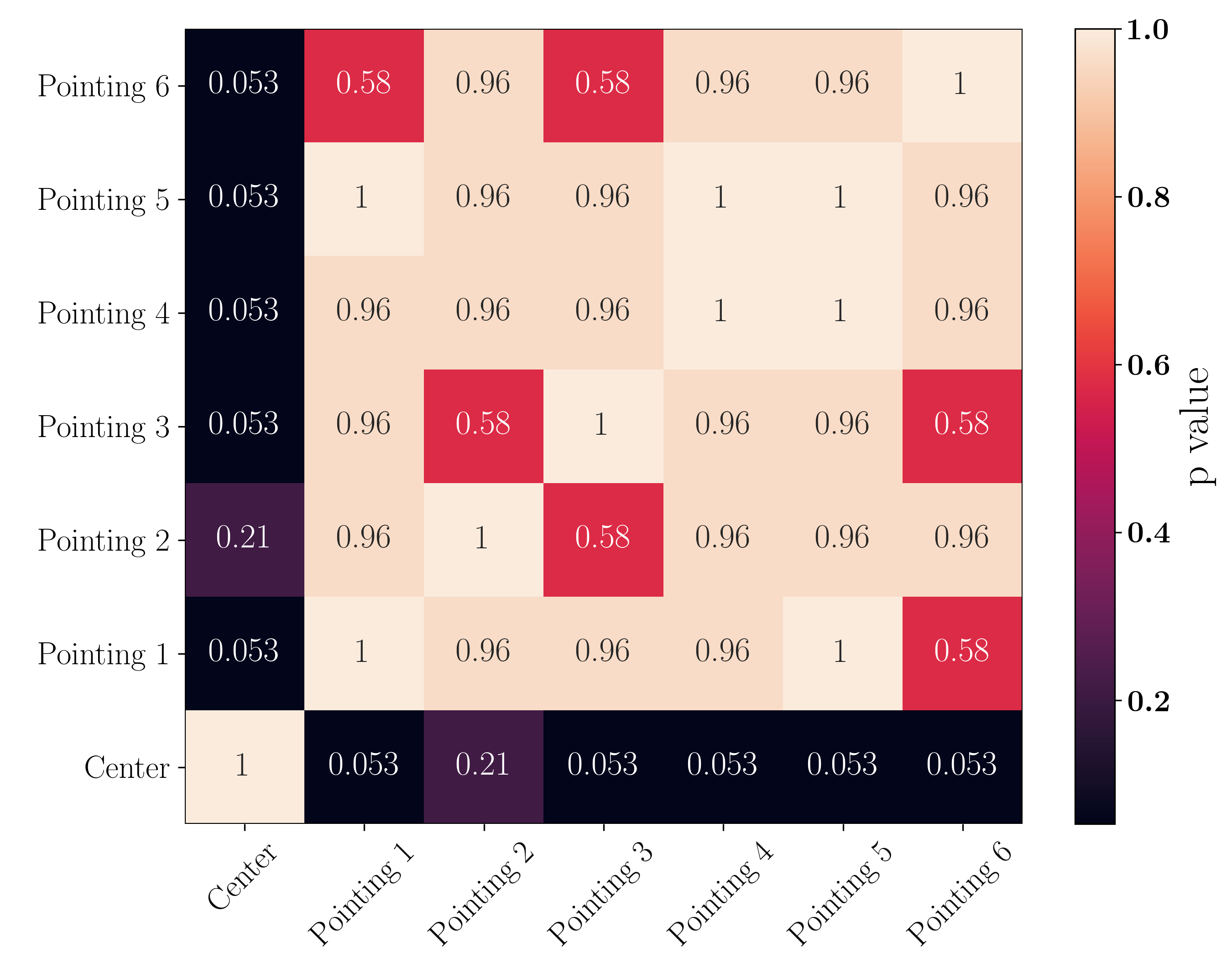}
    \caption{KS statistics from the comparison between the multi-wavelength population distributions of all the pointings.}
    \label{KS_fig}
\end{figure}

The results presented in section \S \ref{xray_ir_counterpart}, on the multi-wavelength population distribution in the cluster, include the analysis of azimuthally averaged radial trends which would require the multi-wavelength distribution of the outer pointings to be similar at a given distance from the reference center. In order to quantify any differences between each individual pointing we performed a KS test on the distribution of radio, X-ray and NIR populations and their correlations. We used the Venn diagram of these three populations within the inner $r=2\arcmin0$ of each pointing (see Figure 7 in \citet{for16}). At this radius there is no overlap between the areas to be compared and, for the radio population, reflects a good compromise between the number of detections and any effect of the primary beam response. Figure \ref{KS_fig} shows the resulting statistics from the KS test where the values correspond to the `p-value' (also color-coded by the same parameter). While the majority of the pointings indicate that their multi-wavelengths distributions are essentially the same or comparable (p-value$\sim$1 or p-value$\gtrsim$6), the central pointing clearly proves the most distinct multi-wavelength population distribution against any of the outer pointings. The similarity shown by the outer pointings allows us to assume that there are no indications for significant differences between their multi-wavelength population distributions and therefore we can safely azimuthally average their radial distributions as a function of projected distance to $\theta^1$ Ori C in order to assess the impact that the most massive stars in the inner ONC has in the multi-wavelength radial distribution of the cluster.

%%%%   Figure B1

% Don't change these lines  FINAL DEFAULT LINES 
\bsp	% typesetting comment
\label{lastpage}
\end{document}